\documentclass[twocolumn,amsmath,amssymb,nofootinbib,superscriptaddress,longbibliography,floatfix]{revtex4-1}

\usepackage{graphicx}
\usepackage{amsmath}
\usepackage{amssymb}
\usepackage{xcolor}
\usepackage{hyperref}
\hypersetup{pdfstartview={FitH},pdfpagemode={UseNone}, breaklinks=true, colorlinks=true, linkcolor=blue, citecolor=blue, urlcolor=blue, bookmarksopen=true, pdfnewwindow=true}
\usepackage[all]{hypcap}
\usepackage[english]{babel}
\usepackage{physics}

\usepackage{array}
\usepackage{tabularx}
\usepackage{multirow}

\usepackage{enumitem}

\newcommand{\sfrac}[2]{#1/#2}

\begin{document}

\preprint{APS/123-QED}

\title{Micellization in active matter of asymmetric self-propelled particles: Experiments}

\author{Anastasia~A.~Molodtsova}
\email{a.molodtsova@metalab.ifmo.ru}
\affiliation{School of Physics and Engineering, ITMO University, 197101 Saint Petersburg, Russian Federation}

\author{Mikhail~K.~Buzakov}
\affiliation{School of Physics and Engineering, ITMO University, 197101 Saint Petersburg, Russian Federation}

\author{Oleg~I.~Burmistrov}
\affiliation{School of Physics and Engineering, ITMO University, 197101 Saint Petersburg, Russian Federation}

\author{Alina~D.~Rozenblit}
\affiliation{School of Physics and Engineering, ITMO University, 197101 Saint Petersburg, Russian Federation}

\author{Vyacheslav~A.~Smirnov}
\affiliation{School of Physics and Engineering, ITMO University, 197101 Saint Petersburg, Russian Federation}

\author{Daria~V.~Sennikova}
\affiliation{School of Physics and Engineering, ITMO University, 197101 Saint Petersburg, Russian Federation}

\author{Vadim~A.~Porvatov}
\affiliation{School of Physics and Engineering, ITMO University, 197101 Saint Petersburg, Russian Federation}

\author{Ekaterina~M.~Puhtina}
\affiliation{School of Physics and Engineering, ITMO University, 197101 Saint Petersburg, Russian Federation}

\author{Alexey~A.~Dmitriev}
\affiliation{School of Physics and Engineering, ITMO University, 197101 Saint Petersburg, Russian Federation}

\author{Nikita~A.~Olekhno}
\email{nikita.olekhno@metalab.ifmo.ru}
\affiliation{School of Physics and Engineering, ITMO University, 197101 Saint Petersburg, Russian Federation}

\date{\today}

\begin{abstract}
Active matter composed of self-propelled particles features fascinating self-organization phenomena, spanning from motility-induced phase separation to phototaxis to topological excitations depending on the nature and parameters of the system. In the present paper, we consider micelle formation by active particles with a broken symmetry having a circular back and a sharpened nose toward which the particles accelerate. As we demonstrate in experiments with robotic swarms, such particles can either remain in the isotropic phase or form micelles depending on the location of their center of inertia, in accordance with a recent theoretical proposal [T.~Kruglov and A.~Borisov, \textit{Presentations and Videos to 7th Edition of the International Conference on Particle-based Methods} (2021), Vol. CT07, p. 2]. Such a behavior is observed for both nonchiral particles moving linearly and placed in a parabolic potential and for chiral particles moving along circular trajectories on a flat surface. By performing experiments with single robots and two-robot collisions, we unveil that the observed emergence of micellization associated with shifting robots' center of inertia towards their noses is governed by at least two-particle effects, in particular, by a difference in the formation of stable two-robot clusters. Finally, we consider the dependence of micelle lifetime and formation probability as well as two-robot collisions on friction between the lateral surfaces of the robots. Crucially, the predicted micellization does not involve any solvation shells that give rise to the micellization of surfactants but is instead driven by an interplay of activity and particle shape asymmetry.
\end{abstract}

\maketitle

\section{Introduction}
\label{sec:Introduction}

Large assemblies of particles that can self-propel or self-rotate by converting internal or ambient energy resources to directed motion demonstrate the emergence of collective phenomena~\cite{1987_Reynolds, 1995_Vicsek} and are referred to as \textit{active matter}~\cite{2012_Vicsek}. Such systems span the entire range of soft condensed matter, from tissues~\cite{2021_Balasubramaniam} and bacterial colonies~\cite{2018_Pierce, 2020_Beer} to colloidal particles~\cite{2013_Bricard} or even collections of simple moving robots~\cite{2013_Giomi, 2019_Barois, 2020_Patterson, 2021_Li, 2022_Boudet}. There is a rich variety of self-organization phenomena and clustering effects in active matter systems, including motility-induced phase separation~\cite{2015_Cates}, the formation of colloidal crystals~\cite{2013_Palacci, 2018_Ginot, 2019_Mousavi}, the emergence of chiral edge states~\cite{2020_Yang, 2022_Shankar, 2017_Souslov}, and diverse physics of topological defects~\cite{2020_Duclos}, to name a few.

However, the formation of micelles—round or spherical-shaped assemblies of elongated particles in which they orient one of their nonequivalent edges to the inner region of a cluster and the other one to the outer region—has been demonstrated only in Janus particles covered with surfactants~\cite{2015_Zhang}. This phenomenon appeared elusive in self-propelled particles without charge displacement, until the recent theoretical proposal of active - matter micellization driven purely by particle shape asymmetry~\cite{2021_Kruglov} and experimental studies of rotelles, the rotational analogs of micelles formed by surfactant-like chain of self-rotating active particles~\cite{2021_Scholz}.

\begin{figure}[t]
    \centerline{\includegraphics[width=8.5cm]{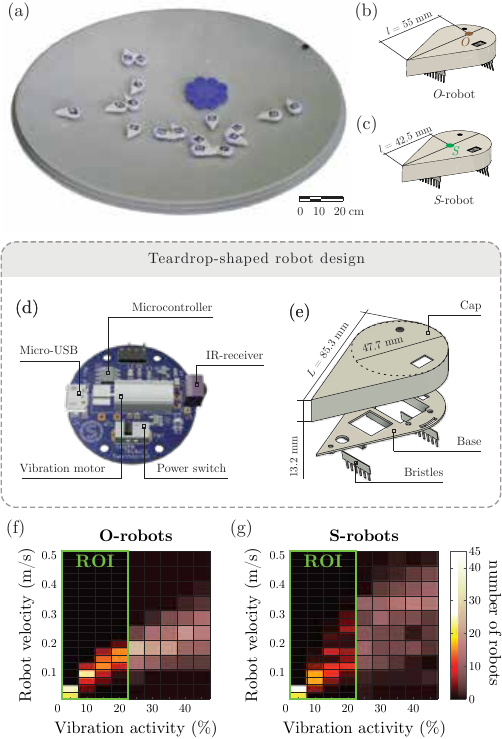}}
    \caption{(a) Robotic swarm in a parabolic potential artistically demonstrating the formation of a single micelle (shaded with purple). The scale bar is $20$~cm. [(b) and (c)] Schematics of the Swarmodroid robot implementing a teardrop-shaped particle with a center of inertia located either at (b) $55.0$~mm from the nose, point $O$ (denoted as $O$-robot) or (c) $42.5$~mm from the nose, point $S$ (referred to as $S$-robot) which are considered in the experiments. (d) Three-dimensional model of the circuit board with an electric motor rotating an unbalanced mass whose vibration is converted to a self-propelled motion of the robot by elastic bristles. (e) Explosion diagram showing the geometry and dimensions of the plastic parts composing a single robot, including the bristles (bottom), the base (middle), and the cap (top). [(f) and (g)] Motion velocity histograms for $O$-robots and $S$-robots at different vibration activities measured as the duty cycle of motor voltage pulse-width modulation. The colorbar shows the number of robots in each bin of the histogram of robot velocities at a given vibration activity. The region of interest (labeled ROI) highlights the activities studied throughout this paper.}
    \label{fig:System}
\end{figure}

In the present paper, we consider ensembles of self-propelled particles that have the teardrop shape shown in Fig.~\ref{fig:System}(a). Particles having a similar shape occur in biological systems, for example, specific types of blood cells~\cite{2014_Gutgemann}, and can be fabricated on the millimeter to micrometer scale with the help of various techniques, e.g., self-propelled asymmetric Janus particles in Refs.~\cite{2016_Dai, 2020_Shah, 2022_Liu}. To demonstrate experimentally that such particles can form micellelike clusters depending on the center of inertia location at the particle axis, we construct a swarm of teardrop-shaped vibrating robots (bristle-bots) shown in Figs.~\ref{fig:System}(b) and~\ref{fig:System}(c) which are based on the Swarmodroid~1.0 platform~\cite{2023_Dmitriev}. Such robots move with a controlled level of activity and are placed in a shallow parabolic potential of a satellite dish to prevent condensation at the system boundary, typical of self-propelled particles~\cite{2013_Giomi, 2018_Deblais, 2021_Boudet}, Fig.~\ref{fig:System}(a), or on a flat surface with their trajectories modified to be slightly chiral (corresponding to robots moving along circular loops). In accordance with the predictions of Ref.~\cite{2021_Kruglov}, we report micelle formation both for swarms on a flat surface and in a parabolic potential and address the dependence of their formation probability on the packing density of robots, their center of inertia location, velocity of their motion, and the friction coefficient between lateral surfaces of robots. Finally, we uncover the microscopic mechanisms responsible for such a sharp dependence of the micellization on the location of individual robots' center of inertia by performing experiments with single robots as well as with two-robot collisions in different conditions.

The paper is organized as follows. In Sec.~\ref{sec:Setup}, we describe our experimental platform. Then, in Sec.~\ref{sec:Micellization_Parameter} a key quantitative parameter is introduced which we use to detect micellization. Section~\ref{sec:Experiments} presents the results of experiments with robotic swarms and includes three subsections. Section~\ref{sec:Condensation} includes studies of the condensation of robots at the boundary when they propel along straight trajectories on a flat surface. The next Sec.~\ref{sec:Chiral_Exp} features a demonstration of robots' micellization on a flat surface for robots moving along chiral trajectories and includes the study of the micellization statistical properties. Then, Sec.~\ref{sec:Parabolic} considers micellization of robots in a parabolic potential, including the dependence of the lifetime and formation probability of micelles on the packing density of robots, their motion velocity, and the friction between their lateral surfaces for different locations of the robots' center of inertia. The experiments considering microscopic origin of the observed micellization transition are discussed in Sec.~\ref{sec:Microscopic}. Section~\ref{sec:Single_Robot} includes studies of individual robot motion in a parabolic potential for different center of inertia locations demonstrating that the behavior at a single-robot level does not demonstrate any pronounced changes. Section~\ref{sec:Two_Robots} features the studies of two-robot collisions pointing toward a considerable increase in the formation of stable quarter-micelle clusters, hinting towards its possible influence on the emergence of the micellization. Section~\ref{sec:Discussion} contains the final remarks and discussion of the obtained results.

\begin{figure*}[t]
    \centering
    \includegraphics[width=17cm]{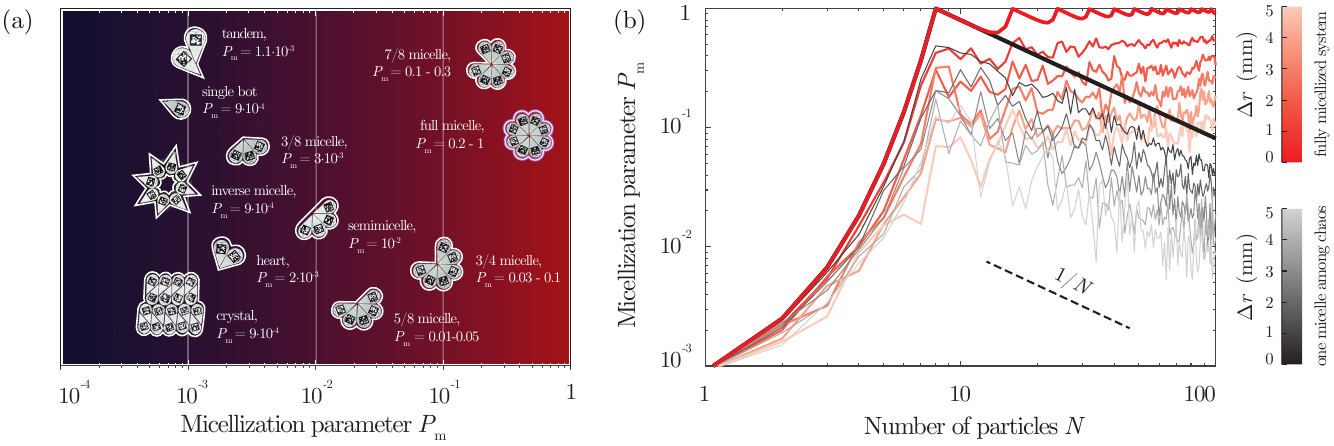}
    \caption{Micellization parameter values for different robotic swarm configurations. (a) Characteristic values of the micellization parameter $P_{\rm m}$ defined by Eq.~\eqref{eq:Parameter} evaluated for several specific types of robot clusters shown in the insets. (b) Numerically calculated dependencies of the micellization parameter $P_{\rm m}$ on the number of particles $N$ for a fully micellized system (the red solid lines) and for a single micelle placed in a set of randomly arranged particles (the black solid lines). The bold curves in panel (b) correspond to idealized systems, while the semitransparent ones are calculated for the systems with additional random shift vectors added to the particle positions, defined by normally distributed components with the zero mean and the dispersion $\Delta r$ in the range $1..5$~mm. The dashed black line at the bottom illustrates power-law asymptotics $1/N$.}
    \label{fig:OrderParam}
\end{figure*}

\section{Experimental platform for active matter: asymmetric robots}
\label{sec:Setup}

In our experiments, we implement the teardrop-shaped particles as self-propelled bristle-bots converting their vibration to a directed motion, Figs.~\ref{fig:System}(b)-~\ref{fig:System}(e). Such robots consist of a printed circuit board [Fig.~\ref{fig:System}(d)] carrying a vibration motor, a battery, and circuitry for infrared remote control, which allows turning the robots on and off simultaneously and varying their vibration activity (i.e., a self-propulsion velocity) by changing the pulse width modulation (PWM) duty cycle of the motor voltage, and a three-dimensional (3D) printed body with elastic bristles at the bottom [Fig.~\ref{fig:System}(e)]~\cite{2023_Dmitriev}. The system dynamics is extracted using a tracking pipeline based on our custom recognition software~\cite{2023_Dmitriev} and the use of ArUco markers.

The length of each robot is $L = 85.3~\text{mm}$, the diameter of the round part is $47.7~\text{mm}$, and the nose angle is $45^{\circ}$ corresponding to $M = 8$ particles in a complete micelle. We choose such a value of $M$ to clearly distinguish micellization from crystallization, as the $C_{8}$ point symmetry is incompatible with a crystalline order. The height of the robots including the bristles is $26$~mm. The robot's center of inertia lies on its mirror symmetry axis at the distance $l$ from its nose cusp. With the help of an additional load, we are able to move the center of inertia from $l=55.0 \pm 1.0$~mm (point $O$), which is the center of inertia of a robot without additional load [Fig.~\ref{fig:System}(b)], to $l=42.5 \pm 1.0$~mm (point $S$) for a robot with a load [Fig.~\ref{fig:System}(c)]. We attach a load consisting of an M6$\times$16 screw and three M6 nuts made of steel, which are fastened to a hole near the robot's nose, to move the center of inertia to point $S$. In the following, we denote such robots as $S$-robots. To increase the mass of a robot by the same amount while maintaining the center of inertia at point $O$, we instead add a load consisting of a steel DIN125A M18 washer, which is glued to the robot cap at point $O$~\cite{Supplement}. We denote these robots as $O$-robots. The mass of $O$-robots and $S$-robots is the same and equal to $30.8\pm 0.2$~g. We study swarms composed of $N = 15$, $30$, $45$, and $46$ robots.

Figures~\ref{fig:System}(f) and~\ref{fig:System}(g) show the velocities of $O$- and $S$-robots, respectively, as dependencies of the PWM duty cycle, which represents the vibration activity. Due to the complex dynamics involved in the conversion of vibration to directional motion by bristles~\cite{2013_Giomi, 2016_Scholz, 2021_Porvatov, 2023_Dmitriev}, there is a distribution of velocities between different robots at each value of the vibration activity, which we show as a color-encoded histogram. The standard deviation of this distribution reaches one third of the mean value for $S$-robots at the vibration activity of $\text{PWM} = 50\%$. Therefore, we do not refer directly to the robot velocity throughout the paper, but instead use vibration activity which is exactly reproducible. It is seen that, despite their different weight distributions, the velocities of $O$-robots and $S$-robots coincide for vibration activities up to $\text{PWM} = 20\%$ that are studied in the following.

\section{Micellization parameter}
\label{sec:Micellization_Parameter}

To quantify the formation of micelles, we introduce the micellization parameter
\begin{equation}
    P_{\rm m} = \frac{1}{N}\sum_{i=1}^{N}{\rm exp}\left(-\left|\sum_{j = 1}^N{\rm e}^{-\sfrac{|\mathbf{r}_{\rm i} - \mathbf{r}_{\rm j}|}{\lambda}} - M\right|\right),
    \label{eq:Parameter}
\end{equation}
where $\mathbf{r}_{\rm i}$ and $\mathbf{r}_{\rm j}$ are the in-plane coordinates of the noses of particles with indices $i$ and $j$, $M$ is the number of particles in a complete micelle (in our case, $M=8$), $\lambda$ is the characteristic distance between the noses of particles at which they are considered forming a micelle (here, we choose $\lambda = L/5 = 17.06$~mm), and $N$ is the total number of particles in the system. Figure~\ref{fig:OrderParam}(a) shows the values of such micellization parameter extracted experimentally for several model systems: an isolated complete micelle with $N=M=8$, several types of isolated incomplete micelles with $N=\{3,4,5,6,7\}$, an inverse micelle with $N=8$, a structure of $N=18$ densely packed robots with crystalline order, a single robot, and two types of two-robot clusters (``tandem'' and ``heart'') that, as will be shown in Sec.~\ref{sec:Microscopic}, are important for understanding the processes facilitating micelle formation. The terms in the sum by index $i$ tend to unity for a complete micelle while rendering values orders of magnitude smaller for other configurations. However, due to the exponential dependence on the distance between the robots' noses, the micellization parameter yields values somewhat lower than unity for experimentally observed micelles, with the majority of observed micelles corresponding to $P_{\rm m} = 0.2 \ldots 0.4$. For incomplete micelles and nonmicellar structures, the micellization parameter $P_{\rm m}$ yields exponentially low values. The normalization prefactor $1/N$ ensures the invariance of the parameter with respect to the number of robots. For fully chaotic systems, the micellization parameter yields the value $P_{\rm m} = e^{-7} \simeq 9 \cdot 10^{-4}$ independently of the number of robots in the system. The same value corresponds to the system consisting of a single robot and is the lowest possible value of the micellization parameter.

\begin{figure*}[t]
    \centerline{\includegraphics[width=17cm]{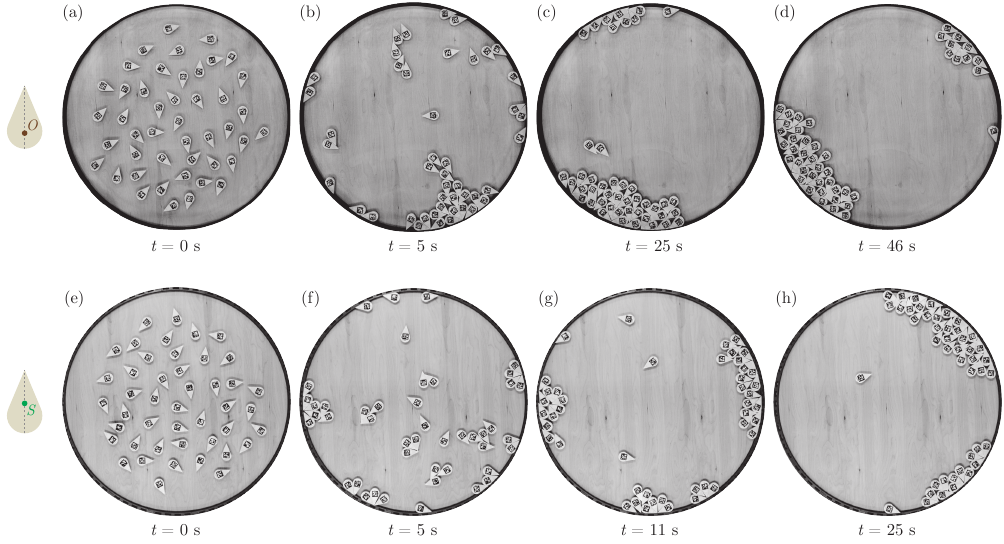}}
    \caption{Condensation of the robotic swarms at the circular boundary with the diameter $D = 90$~cm and an abrupt profile. Panels (a)-(d) and (e)-(h) demonstrate the snapshots of the swarm motion for two realizations featuring different center of inertia locations corresponding to $O$-robots (top row) and $S$-robots (bottom row), starting from different initial arrangements of $N=45$ robots moving at high activity ($\text{PWM}=20\%$). Snapshot times are indicated in the respective panels.}
    \label{fig:Condensation}
\end{figure*}

Figure~\ref{fig:OrderParam}(b) shows the numerically calculated dependencies of the micellization parameter on the number of particles $N$ in the system up to $N=100$. The fully micellized systems [the solid red lines in Fig.~\ref{fig:OrderParam}(b)] correspond to the particles that are added to the system sequentially to form complete micelles. In this case, the micellization parameter is at its maximum for the numbers of particles that are multiples of eight, with intermediate values corresponding to the presence of incomplete micelles resulting in the lower values of $P_{\rm m}$. The curves for a single micelle among a disordered environment [the black solid lines in Fig.~\ref{fig:OrderParam}(b)] are obtained by adding randomly oriented particles at random coordinates within a $80L \times 80L$ box around a complete micelle. The coordinates and angular orientations of the added particles are distributed uniformly. In this case, the micellization parameter decreases as $1/N$ with the number of particles. Therefore, the parameter introduced in Eq.~\eqref{eq:Parameter} characterizes the fraction of micelles in the swarm.

We also study the dependence of the micellization parameter on the distance between the particles in a single micelle by introducing random shifts of the particles' noses from the center of a micelle, Fig.~\ref{fig:OrderParam}(b). The components of the shift vectors are normally distributed with the zero mean and the dispersion $\Delta r$ that varies from $0$ to $5$~mm (approximately $1/10$ of the robot's diameter). The micellization parameter decreases exponentially depending on this amplitude, producing the same value $P_{\rm m} \approx 0.1$ for a complete micelle with $5$~mm gaps between the robots and an ideal incomplete micelle without one robot.

\section{Experimental studies of micellization in robotic swarms}
\label{sec:Experiments}

\subsection{Boundary condensation of non-chiral robots on a flat surface surrounded by a vertical wall}
\label{sec:Condensation}

\setlength\tabcolsep{0.09cm}
\begin{table}[b]
    \renewcommand{\arraystretch}{1.3}
    \centering
    \begin{tabular}{c c c c c c c c c c c c c}
        \hline\hline

        \multicolumn{13}{c}{$O$-robots} 
        \\\hline

        \textnumero & 1 & 2 & 3 & 4 & 5 & 6 & 7 & 8 & 9 & 10 & Avr,~s & Msd,~s
        \\\hline

        \textbf{\textit{t},~s} & \textbf{83} & \textbf{76} & \textbf{83} & \textbf{69} & \textbf{21} & \textbf{17} & \textbf{10} & \textbf{52} & \textbf{24} & \textbf{70} & \multirow{3}{*}{\textbf{49.6}} & \multirow{3}{*}{\textbf{24.6}}
        \\\cline{1-11}

        \textnumero & 11 & 12 & 13 & 14 & 15 & 16 & 17 & 18 & 19 & 20 &  & 
        \\\cline{1-11}

        \textbf{\textit{t},~s} & \textbf{74} & \textbf{48} & \textbf{29} & \textbf{47} & \textbf{46} & \textbf{14} & \textbf{24} & \textbf{69} & \textbf{79} & \textbf{57} & &
        \\\hline\hline

        \multicolumn{13}{c}{$S$-robots} 
        \\\hline

       \textnumero & 1 & 2 & 3 & 4 & 5 & 6 & 7 & 8 & 9 & 10 & Avr,~s & Msd,~s
        \\\hline

         \textbf{\textit{t},~s} & \textbf{22} & \textbf{29} & \textbf{32} & \textbf{35} & \textbf{47} & \textbf{48} & \textbf{12} & \textbf{27} & \textbf{41} & \textbf{27} & \multirow{3}{*}{\textbf{33.3}} & \multirow{3}{*}{\textbf{12.9}}
        \\\cline{1-11}

        \textnumero & 11 & 12 & 13 & 14 & 15 & 16 & 17 & 18 & 19 & 20 &  &  
        \\\cline{1-11}

         \textbf{\textit{t},~s} & \textbf{37} & \textbf{18} & \textbf{30} & \textbf{26} & \textbf{32} & \textbf{25} & \textbf{63} & \textbf{26} & \textbf{62} & \textbf{27} & &
        \\\hline\hline
    \end{tabular}
    \caption{Condensation times for swarms of $N = 45$ non-chiral self-propelled robots moving at high activity ($\text{PWM}=20\%$) on a flat surface bounded by a circular barrier with the diameter $D=90$~cm. The results for $O$-robots and $S$-robots are shown in the upper and lower parts of the table, respectively. The average condensation times (Avr) and their mean square deviations (Msd) obtained by averaging over $20$ experiments are indicated to the right, with particular experimental realizations enumerated in the row with the ``\textnumero'' symbol.}
    \label{tab:Condensation}
\end{table}

We start with considering a swarm of $N=45$ teardrop-shaped robots moving on a flat surface and confined by a circular barrier (Table~\ref{tab:Condensation}). The barrier is 3D printed with PLA plastic, and its diameter is $D=90$~cm; see Fig.~\ref{fig:Condensation}. As seen in Figs.~\ref{fig:Condensation}(a)-\ref{fig:Condensation}(d) and Fig.~\ref{fig:Condensation}(e)-\ref{fig:Condensation}(h), the robots demonstrate a pronounced condensation at the barrier for both considered center of inertia locations, with characteristic condensation times $t_{\rm c} \sim 39.3 \pm 19.6$~s for $O$-robots and $t_{\rm c} \sim 33.3 \pm 12.9$~s for $S$-robots, respectively, see supplementary video~1. These values are obtained by averaging over $20$ experiments for each type of the robots starting from a randomly arranged swarm. The detailed results are provided in Table~\ref{tab:Condensation}. In experiments, the measured value of condensation time corresponds to the time frame when all $N$ robots localize at the barrier for the first time, with the only allowed exception of one or two robots roaming along closed circular trajectories in the bulk due to the chirality that is inevitably present in their motion.

Once condensed at the boundary, the robots cannot return to the center, which hinders the further emergence of large clusters in the bulk. Moreover, the robots located at the barrier touch it with their noses or lateral surfaces; hence, an axially symmetric micelle cannot form at the barrier. Similar condensation has already been reported~\cite{2013_Giomi} and even applied to engineer specific swarm behavior~\cite{2018_Deblais} or collective functionalities~\cite{2021_Boudet}. Its appearance appears absolutely intuitive and is related to the negative curvature of the boundary. As shown in Secs.~\ref{sec:Chiral_Exp} and \ref{sec:Parabolic}, the characteristic times of a single micelle formation are much longer than the condensation times observed in Fig.~\ref{fig:Condensation}.

As a result, in order to study micellization, such boundary-assisted condensation of robots should be avoided for as long as is feasible, which can be achieved with several strategies.

\begin{enumerate}[wide, labelwidth=0pt, labelindent=0pt]

\item Considering abrupt barriers of the same type, but with diameters large enough for the system to micellize prior to the condensation is achieved. Indeed, the characteristic time of the bulk micellization transition should be independent of the systems' boundary, while the condensation time grows with the barrier diameter (considering fixed filling density, since it is defined by the mean distance each particle should travel to collide with the barrier).

\item Repulsive boundaries, for example, creating air flows in a similar manner to those described in Ref.~\cite{2021_Li} or incorporating mechanically moving parts to orient the robots' noses back to the center of the experimental area.

\item Passive barriers with specific shapes that facilitate returning robots to the bulk after traveling a certain distance along the boundary. For example, Refs.~\cite{2010_Deseigne,2014_Kumar} consider sunflower-shaped barriers used together with systems of self-propelled disks or millimeter-sized tapered rods, respectively.

\item Chiral trajectories of robots. If each robot moves along closed circular trajectories with diameter $d$ much larger than the characteristic robot size $L \ll d$, yet much lower than the barrier diameter $d \ll D$, then the robots can be considered (locally) self-propelling forward, yet they can localize in the bulk and touch the barrier only as a result of collisions with other robots. Moreover, the robot can still return to the bulk after condensing at the barrier if its motion direction along the barrier coincides with its own chirality. Such chiral systems are considered in Refs.~\cite{2018_Deblais,2020_Barois,2021_Boudet,2023_Siebers}.

\item Soft confining potentials preventing condensation. Robotic swarms in external potentials have been considered in the case of an inclined plane with phototactic behavior of robots whose activity is defined by the level of external illumination~\cite{2023_Zion} and even in more complex cases of robots that can sense and alter the illumination pattern on the surface they travel on~\cite{2021_Wang}.
\end{enumerate}

\begin{figure*}[t]
    \centerline{\includegraphics[width=17cm]{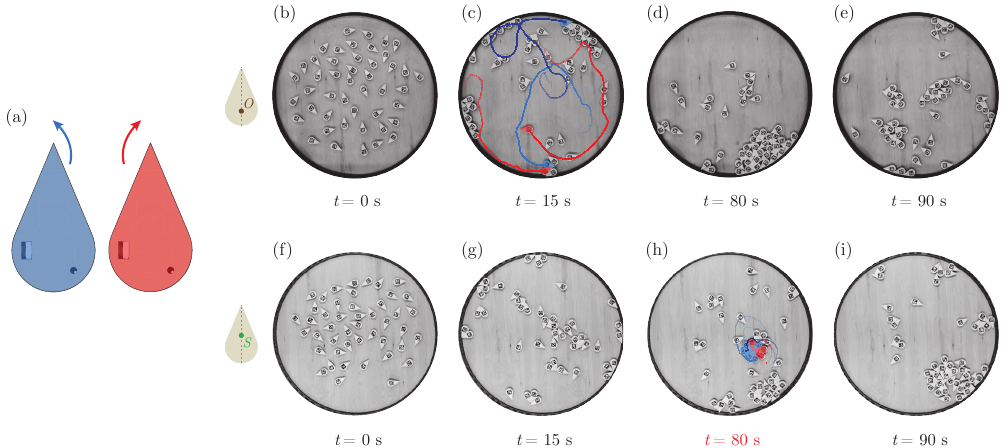}}
    \caption{Micellization in chiral swarms with $N=46$ robots moving on a flat surface bounded by a circular barrier with the diameter $D=90$~cm. (a) Artistic illustration of counter-clockwise (blue) and clockwise (red) chiral robots. (b)-(e) Snapshots of a single experiment with $O$-robots for the timestamps (b) $t=0$~s, (c) $t=15$~s, (d) $t=80$~s, and (e) $t=90$~s counting from the moment the robots are turned on and start moving at high activity ($\text{PWM}=20\%$). Panel (c) also shows several trajectories of CW- and CCW-chiral robots colored accordingly definitions of Panel (a) for the past $15$~seconds from $t=0$~s to $t=15$~s. (f)-(i) The same as Panels (b)-(e), but for a single experiment with $S$-robots and high activity ($\text{PWM}=20\%$). Panel (h) demonstrates the formation of a micelle consisting of three CCW- and five CW-chiral robots. Colored lines in Panel (h) illustrate trajectories of several robots for the past ten seconds from $t=70$~s to $t=80$~s.}
    \label{fig:Chiral}
\end{figure*}

\begin{table*}[tbp]
    \renewcommand{\arraystretch}{1.3}
    \centering
    \begin{tabular} { 
   m{0.5cm} m{0.5cm} m{3cm} m{3cm} m{0.5cm} m{0.5cm} m{0.5cm} m{0.5cm} m{3cm} m{3cm} m{0.5cm} }
        \hline\hline

        \textnumero & 1 & \centering{2} & \centering{3} & 4 & 5 & 6 & 7 & \centering{8} & \centering{9} & 10
        \\\hline

        $O$ & $\times$ & \centering$\times$ & \centering$\times$ & $\times$ & $\times$ & $\times$ & $\times$ & \centering$\times$ & \centering$\times$ & $\times$
         \\

        $S$ & $\times$ & \centering{173--174~s} & \centering{7--8~s} & $\times$ & $\times$ & $\times$ & $\times$ & \centering{174--176~s} & \centering{54--55~s} & $\times$
        \\
         & & \centering{\includegraphics[width=2.5cm]{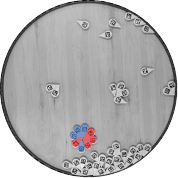}} & \centering{\includegraphics[width=2.5cm]{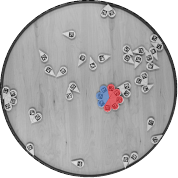}} & & & & & \centering{\includegraphics[width=2.5cm]{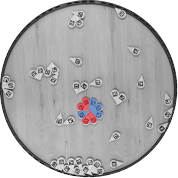}} & \centering{\includegraphics[width=2.5cm]{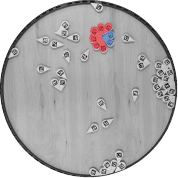}} & 
         \\\hline\hline

         \textnumero & 11 & \centering{12} & \centering{13} & 14 & 15 & 16 & 17 & \centering{18} & \centering{19} & 20
        \\\hline

        $O$ & $\times$ & \centering$\times$ & \centering{$\times$} & $\times$ & $\times$ & $\times$ & $\times$ & \centering$\times$ & \centering$\times$ & $\times$
         \\
        
         $S$ & $\times$ & \centering$\times$ & \centering{22--23~s} & $\times$ & $\times$ & $\times$ & $\times$ & \centering{48--49~s} & \centering{80--83~s} & $\times$
        \\
         & & & \centering{\includegraphics[width=2.5cm]{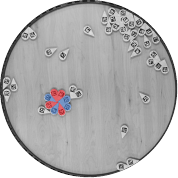}} & & & & & \centering{\includegraphics[width=2.5cm]{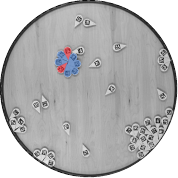}} & \centering{\includegraphics[width=2.5cm]{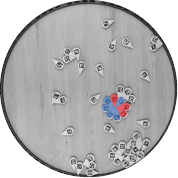}} & 
         \\
          &  &  &  &  &  &  &  & \centering{251--252~s} & & 
        \\
         & & & & & & & & \centering{\includegraphics[width=2.5cm]{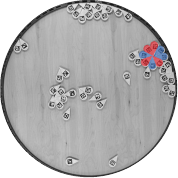}} & & 
         \\\hline\hline
    \end{tabular}
    \caption{Micellization times for $N=46$ chiral robots moving with high activity ($\text{PWM}=20\%$) on a flat surface bounded by a circular barrier with the diameter $D=90$~cm. The robotic swarms with the individual robots' centers of inertia at points $O$ and $S$ are shown in the top and bottom rows under the number of experiment \textnumero, respectively. If micelles are formed during a certain experiment, then the timestamps of micelle formation and decomposition are provided in the corresponding cell along with a snapshot from the experiment recording which illustrates a composition and location of the micelles. CW- and CCW-chiral robots are shaded with red and blue respectively, following the notation of Fig.~\ref{fig:Chiral}(b). If no micelle formation is observed during the experiment, then the corresponding cell contains a ``$\times$'' symbol.}
    \label{tab:Chiral_micelle_time}
\end{table*}

In the following, we consider two strategies: chiral robots moving along circular trajectories in clockwise or counterclockwise directions that allow us to demonstrate the emergence of micellization in robotic swarms on flat surfaces, and non-chiral robots in a parabolic potential.

\subsection{Micelle-like clusters in swarms of chiral robots on a flat surface}
\label{sec:Chiral_Exp}

We start with considering swarms with an even number of robots $N = 46$ in which half of the robots have clockwise (CW) chirality, i.e., move along the circular trajectories centered to the right from the robot's propulsion vector direction, while the rest half robots move in the opposite, counterclockwise (CCW) direction, Fig.~\ref{fig:Chiral}(a). The robots are made chiral by adding an additional $10^\circ$ inclination of the bristles in the planes of their arrays. We select the opposite additional inclination angles for the front and bottom bristle arrays, thus adding the rotational component to the robots' propulsion. In experiments, we set the number of robots rotating clockwise and counterclockwise equal to zero out the net chirality of the system and avoid chirality-assisted effects such as the formation of unidirectional boundary flows~\cite{2020_Yang,2020_Barois}. As outlined in Sec.~\ref{sec:Condensation}, the robots' trajectories are tuned in such a way that the radius of the trajectory along which the robot moves is less than the radius of the barrier, but is larger than the size of the robot, which allows us to consider the robot's motion locally linear. Moreover, the chirality of the robot's trajectory allows it to turn away from the barrier in certain situations, and thus additionally prevents the condensation mentioned in Sec.~\ref{sec:Condensation}. The considered chiral robots are placed on a flat surface made of plywood and are bounded by a plastic barrier with diameter $D = 90$~cm.

In our experiments, we vary the location of the center of inertia considering two cases: $O$-robots [Figs.~\ref{fig:Chiral}(b)-ref{fig:Chiral}(e)] and $S$-robots [Figs.~\ref{fig:Chiral}(f)-\ref{fig:Chiral}(i)]. We capture the collective dynamics of the robotic swarm on video for five minutes and repeat the experiment $20$~times for each robot type, considering different randomly chosen initial locations of the robots. In all experiments, we set the high robot activity ($\text{PWM}=20\%$).

\begin{figure}[tbp]
    \centering
    \includegraphics[width=8.5cm]{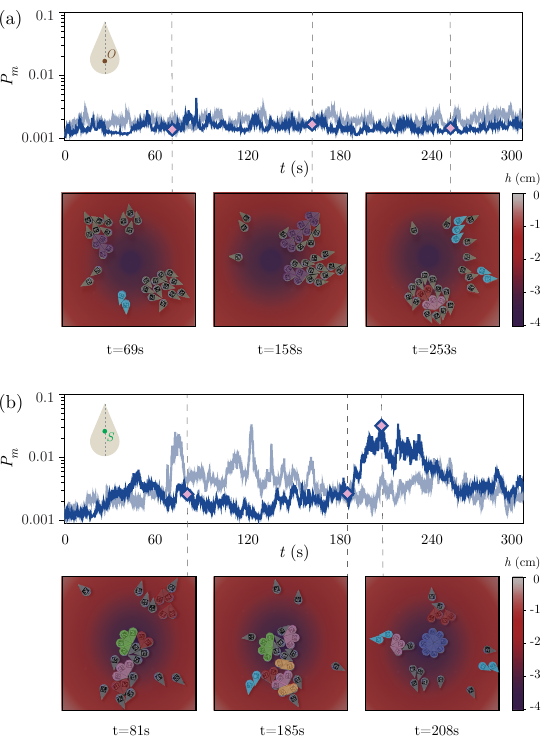}
    \caption{(a) Time dependence of the micellization parameter $P_{m}$ defined by Eq.~\eqref{eq:Parameter} for swarms of $N=30$ $O$-robots moving with low activity ($\text{PWM=10\%}$) and with smooth lateral surfaces (the gray solid line) and with abrasive lateral surfaces (the blue solid line). (b) The same as (a), but for $S$-robots. Dashed vertical lines in both panels mark the timestamps corresponding to the system snapshots shown in the insets to the bottom of each panel and indicated under the corresponding inset. Background color in the insets shows the local height $h$ of the satellite dish defining the parabolic potential at the respective points according to the colorbars shown to the right.}
    \label{fig:Configurations}
\end{figure}

As seen in Figs.~\ref{fig:Chiral}(b)-\ref{fig:Chiral}(e), for $O$-robots micelle formation is not observed, although in Fig.~\ref{fig:Chiral}(c) it is clearly seen that the robots turn away from the boundaries of the barrier without condensing on it (see the red and blue tracks highlighting the trajectories of CW- and CCW-chiral robots, respectively). In contrast, for $S$-robots, a micelle is formed at $t \approx 80$~s, Fig.~\ref{fig:Chiral}(h). However, it decomposes in several seconds after the formation and is not present in Fig.~\ref{fig:Chiral}(i), suggesting general instability of micelles in the system with chiral robots. Both cases are illustrated in supplementary video~2.

The reason behind such an instability is the following. In a two-robot cluster (representing a quarter of a micelle) composed of a CW-chiral robot placed to the left from a CCW-chiral robot (looking towards the direction of its propulsion vector), two such robots possess propulsion components that attract them together while they propel forward in the same direction. As a result, such a cluster retains its shape during motion. The formation of this type of two-robot cluster is analyzed in more detail in Sec.~\ref{sec:Two_Robots}. However, if we instead put the CW-robot to the right of the CCW-robot, then repulsive propulsion components appear that pull the robots apart. Therefore, such a cluster cannot exist for a long time. The stability of analogous configuration with two CW- or two CCW-chiral robots will, obviously, be intermediate between the two considered cases. The same logic applies to configurations with larger numbers of robots, i.e., half-micelles, complete micelles, and other clusters. In addition, every cluster can be composed with several different combinations of CW- and CCW-chiral robots, for example, the micelle in Fig.~\ref{fig:Chiral}(h) is composed from three CW- and five CCW-chiral robots. The number of these combinations increases with the number of robots in the cluster, as is known from combinatorics, and only part of these configurations are stable.

Next, we analyze the statistical properties of the micelles observed in the discussed experiments with swarms of chiral robots on a flat surface. As seen in Table~\ref{tab:Chiral_micelle_time}, no micelles were observed in the system with $O$-robots. However, in the case of $S$-robots, a micelle is formed in $35\%$ of the experiments, with the mean lifetime $t \approx 1.4$~s limited by the instability of the micelles formed by chiral robots. The insets of Table~\ref{tab:Chiral_micelle_time} highlight CW-chiral robots with red and CCW-chiral robots with blue, showing different ratios of CW- and CCW-chiral robots in the observed micelles. Finally, in accordance with our initial considerations, the characteristic time of micelle formation $t_M=101 \pm 87$~s significantly exceeds the typical condensation time $t_c=33.3 \pm 12.9$~s (for $S$-robots, see Sec.~\ref{sec:Condensation}) significantly, justifying the necessity of considering chiral trajectories.

To summarize, condensation at the vertical wall barrier is significantly reduced in swarms of chiral robots moving on a flat surface compared to nonchiral robots. Moreover, micelle formation is observed only in the case of $S$-robots. However, a quantitative analysis of the micellization is complicated by the short micelle lifetime of just a few seconds, making it necessary to consider various strategies toward its increase.

\subsection{Micellelike clusters in swarms of nonchiral robots in a parabolic potential}
\label{sec:Parabolic}

We proceed with studying a system of $N=45$ nonchiral $O$-robots moving with low activity ($\text{PWM} = 10\%$)[Fig.~\ref{fig:Configurations}(a)] and placed in a parabolic satellite dish having the dimensions of $120 \times 110 \times 11~{\rm cm}^{3}$ [Fig.~\ref{fig:System}(a)]. In contrast to a hard wall, such a dish creates a soft localizing potential and, as we will demonstrate further, prevents the condensation of robots at the boundary. The micellization parameter $P_{\rm m}$ fluctuates during the system evolution but does not demonstrate any pronounced growth characteristic of micellization, Fig.~\ref{fig:Configurations}(a). In this case, robots are predominantly directed away from the vertex of the parabolic potential~\cite{2019_Dauchot}. Covering the lateral surfaces of robots with abrasive tape in order to increase the friction between them does not result in any systematic changes of the micellization parameter, see the blue solid line in Fig.~\ref{fig:Configurations}(a).

The situation changes qualitatively for $S$-robots, Fig.~\ref{fig:Configurations}(b). Initially randomized, the system first demonstrates a low micellization parameter corresponding to the absence of micelles (see the inset for $t = 81~\text{s}$ showing the robots' distribution), followed by an increase in the micellization parameter corresponding to formation of incomplete micelles (see the inset for $t = 185~\text{s}$). Finally, at $t = 200 \ldots 230~\text{s}$, a plateau of the micellization parameter is observed, corresponding to the formation of a stable micelle shown in the inset at $t = 208~\text{s}$, as demonstrated in \href{https://github.com/swarmtronics/supplementary-files/raw/refs/heads/main/Micellization_in_active_matter_of_asymmetric_self-propelled_particles_-_Experiments/SUPPLEMENTARY%20VIDEO%203.mp4}{Supplementary Video~3}.

\begin{figure*}[tbp]
    \centering
    \includegraphics[width=\textwidth]{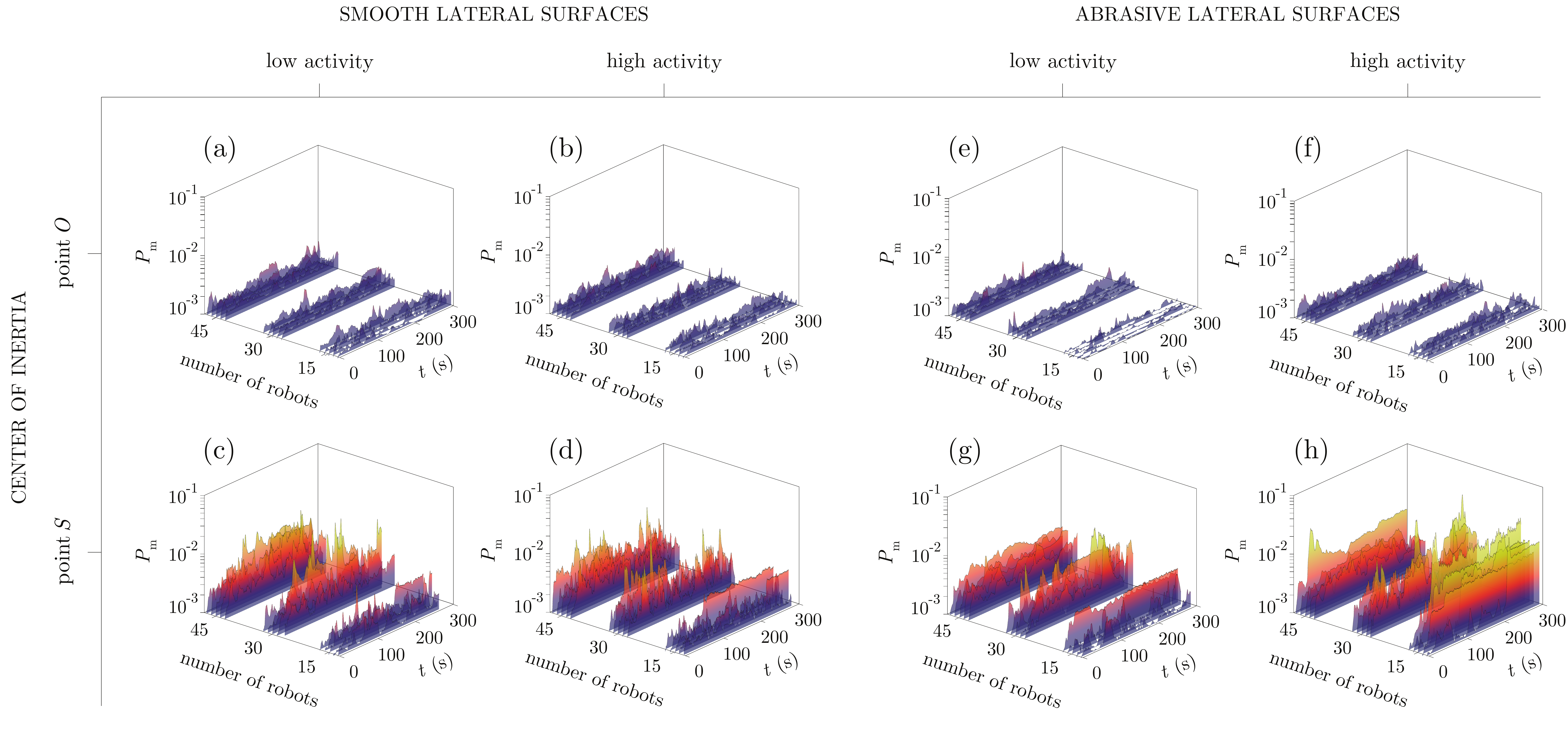}
    \caption{Experimental studies of the micellization parameter $P_{\rm m}$ defined by Eq.~\eqref{eq:Parameter} as a function of time $t$ in swarms of teardrop-shaped particles. (a) Experiments for the setup with $O$-robots with smooth lateral surfaces moving at low activity (PWM = $10\%$). Plots are grouped by numbers of robots $N=15$, $N=30$, and $N=45$. Five experimental realizations are shown for each $N$, corresponding to different random initial positions of the robots. (b) The same as panel (a), but for the increased robot activity (PWM = $20\%$). [(c) and (d)] The same as in (a) and (b), but for $S$-robots. [(e)-(h)] The same as in (a)-(d), but for the robots with lateral surfaces covered by an abrasive.}
    \label{fig:Experiments_Tabl}
\end{figure*}

Next, we consider the dependencies of the micellization parameter on the activity of the robots, their center of inertia location, and the packing density that is varied by changing the number of robots $N$ in the parabolic potential. For systems of $O$-robots with low ($\text{PWM} = 10\%$) and high ($\text{PWM} = 20\%$) activity shown in Figs.~\ref{fig:Experiments_Tabl}(a) and \ref{fig:Experiments_Tabl}(b) the micellization parameter fluctuates near the minimal level ($P_{\rm m} \simeq 9\times 10^{-4}$) independently of the density of the system, and micelle formation is not observed; see Table~\ref{tab:Potential_micelle_time}. $S$-robots do not demonstrate any micelles as well in the experiments with sparse swarms of $N=15$ robots; see Figs.~\ref{fig:Experiments_Tabl}(c) and \ref{fig:Experiments_Tabl}(d). However, stable half-micelles appear in these systems, characterized by an increase in the micellization parameter ($P_{\rm m} \simeq 4\times 10^{-3}$). The formation of transient micelles with a short lifetime of a few seconds emerges at the higher density $N=30$, and is well expressed for $N=45$, indicated by multiple peaks of the micellization parameter in Figs.~\ref{fig:Experiments_Tabl}(c) and \ref{fig:Experiments_Tabl}(d) both for low ($\text{PWM}=10\%$) and high ($\text{PWM}=20\%$) activities. Increasing robot activity results in shorter average micelle lifetimes, namely $2.7$~s for $N = 30$ and $3.8$~s for $N = 45$ for high robot activity, compared to $5.7$~s for $N = 30$ and $10.1$~s for $N = 45$ for low activity.

Next, we modify the system by covering the lateral surfaces of all robots with an abrasive (P800 sandpaper)~\cite{Supplement} to consider how the micellization changes in the case of high friction between the particles, which may prove important for microscale implementations. As seen in Figs.~\ref{fig:Experiments_Tabl}(e) and \ref{fig:Experiments_Tabl}(f), the behavior of $O$-robots is similar to the case of particles with low lateral friction; Figs.~\ref{fig:Experiments_Tabl}(a) and \ref{fig:Experiments_Tabl}(b). The micellization parameter slightly fluctuates, indicating the absence of micellization. The same is confirmed by the data in Table~\ref{tab:Potential_micelle_time}.

\setlength\tabcolsep{0.01cm}
\begin{table*}[tbp]
    \renewcommand{\arraystretch}{1.3}
    \centering
    \begin{tabularx}{\textwidth} { 
   >{\centering\arraybackslash}X
   >{\centering\arraybackslash}X
   >{\centering\arraybackslash}X >{\centering\arraybackslash}X >{\centering\arraybackslash}X >{\centering\arraybackslash}X >{\centering\arraybackslash}X >{\centering\arraybackslash}X >{\centering\arraybackslash}X >{\centering\arraybackslash}X >{\centering\arraybackslash}X >{\centering\arraybackslash}X >{\centering\arraybackslash}X }
        \hline\hline
        \multicolumn{13}{c}{Smooth lateral surfaces} 
        \\\hline
        \multicolumn{1}{c}{}
         & 
        \multicolumn{6}{c}{Low activity}
         & 
         \multicolumn{6}{c}{High activity}
         \\\hline
         \multicolumn{1}{c}{}
         &
         \multicolumn{2}{c}{$N=15$}
         &
         \multicolumn{2}{c}{$N=30$}
         &
         \multicolumn{2}{c}{$N=45$}
         &
         \multicolumn{2}{c}{$N=15$}
         &
         \multicolumn{2}{c}{$N=30$}
         &
         \multicolumn{2}{c}{$N=45$}
         \\\hline
         \textnumero & $O$ & $S$ & $O$ & $S$ & $O$ & $S$ & $O$ & $S$ & $O$ & $S$ & $O$ & $S$ 
         \\\hline
         
         1 & $\times$ & $\times$ & $\times$ & 69--77~s & $\times$ & 224--226~s & $\times$ & $\times$ & $\times$ & 70--71~s & $\times$ & 71--75~s 
         \\
          & &  &  & 121--123~s &  &  &  &  &  & 102--106~s &  & 116--121~s
         \\
         & &  &  &  &  &  &  &  &  &  &  & 127--131~s
         \\
         & &  &  &  &  &  &  &  &  &  &  & 229--230~s
         \\
         & &  &  &  &  &  &  &  &  &  &  & 248--251~s
         \\
         & &  &  &  &  &  &  &  &  &  &  & 284--285~s
         \\

         2 & $\times$ & $\times$ & $\times$ & 142--146~s & $\times$ & 106--116~s & $\times$ & $\times$ & $\times$ & 59--60~s & $\times$ & 66--75~s
         \\
         & & & & 186--198~s & & 147--149~s & & & & 222--225~s & & 102--105~s
         \\
         & & & & 212--214~s & & & & & & & & 
         \\
    
         3 & $\times$ & $\times$ & $\times$ & 267--268~s & $\times$ & 184--185~s & $\times$ & $\times$ & $\times$ & $\times$ & $\times$ & $\times$
         \\
          & &  &  & 295--300~s &  &  &  & &  &  &  & 
         \\

         4 & $\times$ & $\times$ & $\times$ & 192--204~s & $\times$ & 157--166~s & $\times$ & $\times$ & $\times$ & 41--43~s & $\times$ & 69--70~s
         \\
         & & & &  & & 189--243~s & & & & 192--193~s &  & 193--194~s 
         \\

          5 & $\times$ & $\times$ & $\times$ & $\times$ & $\times$ & 53--54~s & $\times$ & $\times$ & $\times$ & 41--48~s & $\times$ & $\times$
         \\
         & &  & &  &  & 99--101~s &  &  &  & & &

         \\\hline\hline
         \multicolumn{13}{c}{Abrasive lateral surfaces} 
        \\\hline
        \multicolumn{1}{c}{}
         &
        \multicolumn{6}{c}{Low activity}
         & 
         \multicolumn{6}{c}{High activity}
         \\\hline
         \multicolumn{1}{c}{}
         &
         \multicolumn{2}{c}{$N=15$}
         &
         \multicolumn{2}{c}{$N=30$}
         &
         \multicolumn{2}{c}{$N=45$}
         &
         \multicolumn{2}{c}{$N=15$}
         &
         \multicolumn{2}{c}{$N=30$}
         &
         \multicolumn{2}{c}{$N=45$}
         \\\hline
         \textnumero & $O$ & $S$ & $O$ & $S$ & $O$ & $S$ & $O$ & $S$ & $O$ & $S$ & $O$ & $S$ 
         \\\hline
         
         1 &$\times$ & $\times$ & $\times$ & $\times$ & $\times$ & $\times$ & $\times$ & $\times$ & $\times$ & 169--176~s & $\times$ & 34--47~s 
         \\
         & &  &  &  &  &  &  &  &  & 265--273~s &  &
         \\
         & &  &  &  &  &  &  &  &  & 287--300~s &  &
         \\
         
         2 & $\times$ & $\times$ & $\times$ & 216--232~s & $\times$ & $\times$ & $\times$ & 229--234~s & $\times$ & 41--43~s & $\times$ & $\times$ 
         \\
         & &  &  &  &  &  &  &  &  & 248--252~s &  &
         \\
         
         3 & $\times$ & $\times$ & $\times$ & $\times$ & $\times$ & $\times$ & $\times$ & 217--224~s & $\times$ & 40--44~s & $\times$ & $\times$ 
         \\

         4 & $\times$ & $\times$ & $\times$ & 197--233~s & $\times$ & $\times$ & $\times$ & 62--300~s & $\times$ & 93--94~s & $\times$ & $\times$ 
         \\

         5 & $\times$ & $\times$ & $\times$ & $\times$ & $\times$ & $\times$ & $\times$ & 222--224~s & $\times$ & 234--241~s & $\times$ & $\times$ 
         \\
         & &  &  &  &  &  &  &  &  & 263--270~s &  &

         \\\hline\hline
    \end{tabularx}
    \caption{Micellization time for $N=15$, $N=30$, and $N=45$ self-propelled robots in a parabolic potential for $O$-robots and $S$-robots moving with low activity ($\text{PWM}=10\%$) and high activity ($\text{PWM}=20\%$), and two types of robots' lateral surfaces: abrasive and smooth. In the table, ``$\times$'' denotes the absence of micellization. Two numbers in each cell correspond to the time of micelle formation and the time of its decomposition. If micelles are formed more than one time during a particular experiment, the corresponding cell contains multiple time ranges. Numbers in the column with the ``\textnumero'' symbol enumerate experimental realizations.}
    \label{tab:Potential_micelle_time}
\end{table*}

However, for $S$-robots with abrasive lateral surfaces, the situation becomes strikingly different compared to the case with smooth lateral surfaces. As seen in Table~\ref{tab:Potential_micelle_time}, micelle formation for low activity ($\text{PWM}=10\%$) is observed only for the system with $N=30$, while for sparse swarms with $N=15$ or, in contrast, dense swarms with $N=45$ the micelles are not observed. Moreover, for high activity ($\text{PWM}=20\%$), micelles are observed most frequently for intermediate density $N=30$, while for $N=15$ micelle formation is less pronounced and appears almost completely suppressed for $N=45$, with only a single experiment featuring the formation of a complete micelle. Instead, distorted micelle-like clusters consisting of two half-micelles shifted a few centimeters relative to each other frequently form in these systems, characterized by plateaus of the micellization parameter with values $P_{\rm m} \simeq (5..8)\times 10^{-3}$; see Figs.~\ref{fig:Experiments_Tabl}(g) and \ref{fig:Experiments_Tabl}(h). The average lifetimes of micelles are $4.6$~s for $N=15$ (excluding the anomalously stable micelle discussed separately in Sec.~\ref{sec:Discussion}), $5.7$~s for $N=30$, and $13$~s for $N=45$, significantly exceeding those for the micelles formed by smooth robots. Decreasing robot activity leads to an even longer average micelle lifetime of $26$~s for $N=30$. Thus, it is seen that, in contrast to robots with smooth lateral surfaces, micelles are more stable in swarms of abrasive robots, and increasing robot activity significantly enhances the probability of micelle formation.

This difference in micellization between smooth and abrasive robots could be related to the improved stability of the clusters formed by robots with high lateral friction. Systems of robots with smooth lateral surfaces demonstrate fewer micelles at lower densities, while at higher densities there are more micelles that decompose quickly. This might be explained by the increased probability of micelle formation during the decomposition of other clusters, which are generally characterized by low stability. With high lateral friction, micelles demonstrate longer lifetimes, apparently related to restriction of the relative lateral movement of adjacent robots in a cluster. This would presumably lead to increased micellization in systems forming clusters that are more likely to merge into micelles, which, once formed, would remain stable. In contrast, for systems where disordered clusters are more likely to form, their enhanced stability would lower the chances of decomposition followed by micelle formation, resulting in decreased micellization. Such a variety of possible mechanisms emphasizes the complex interplay between the asymmetry of individual particles, their activity, and friction in the considered system, which may be crucial for microscale realizations, for example, using Janus particles~\cite{2015_Zhang, Hils_2021}.

To evaluate the interaction of micellelike clusters with nonmicellar aggregates, lifetimes of manually assembled micellelike clusters were assessed. First, manually assembled micellelike clusters in the absence of nonmicellized robots were examined. They remained stable for the entire duration of the experiment (several minutes), independently of the vibration activity of the robots (low or high), their lateral friction (smooth or abrasive side surfaces), and even of the center of inertia location-micellelike clusters composed of both $S$-robots and $O$-robots remained stable (see supplementary video~7). From this we can conclude that micellization is governed by the probability of micellelike cluster formation, but once formed, micellelike clusters are stabilized by the parabolic potential. After that, nonmicellized robots were added to evaluate the interaction of micellelike clusters with other aggregates and individual robots. These systems most often displayed two kinds of behavior: micellelike clusters either decomposed in the first few seconds after a collision with surrounding robots, or remained stable for the entire experiment if such collisions did not occur (see supplementary video~8). Additionally, $S$-robots occasionally formed new micellelike clusters, while $O$-robots did not (see the supplementary material~\cite{Supplement} for a detailed analysis). Therefore, micellelike clusters continuously form and decompose with probabilities governed by the interactions between individual robots and various clusters.

\section{Microscopic mechanisms of micelle-like clusters formation}
\label{sec:Microscopic}

To unveil the physical mechanisms behind the increased micellization in swarms of robots with the center of inertia located closer to the robots' noses, we perform a series of experiments with single robots to study how their trajectories vary for different center of inertia locations and consider two-robot collisions to address the changes in their scattering associated with the center of inertia location.

\subsection{Similar single-particle dynamics of robots with different center of inertia locations}
\label{sec:Single_Robot}

\begin{figure*}[p]
    \centering
    \includegraphics[width=\textwidth]{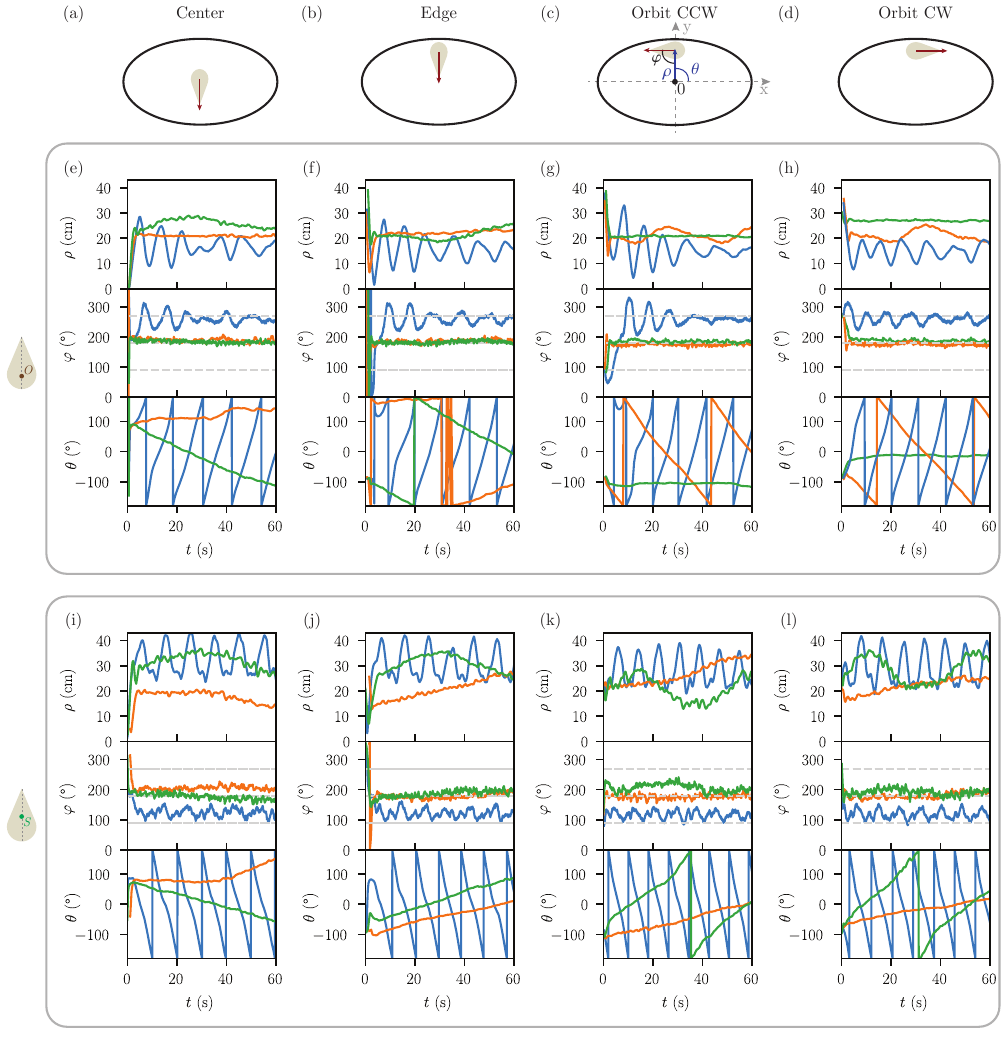}
    \caption{Single robot dynamics in a parabolic potential. [(a)-(d)] Schematic representation of the initial positions and orientations (gray outline) along with the propulsion directions (red arrows) of a single robot in the parabolic satellite dish (black outline). Panel (c) also shows the polar coordinates $\rho$ and $\theta$ used to describe the robot radius vector, as well as its orientation angle $\varphi$. (e) Time dependencies of the radius vector components ($\rho$, $\theta$) and the orientation angle $\varphi$ of an $O$-robot in a parabolic potential. The blue, green, and orange solid lines correspond to three different robots, each moving with high activity ($\text{PWM}=20\%$), starting from the vertex of parabolic potential, as shown in panel (a). The gray dashed lines in the middle panel indicate the angles $90^\circ$, $180^\circ$, and $270^\circ$. [(f)-(h)] The same as panel (e), but for the initial positions and orientations of the robots at the edge of the parabolic potential corresponding to the ones shown in panels (b)-(d): (f) with the robot's propulsion vector directed to the vertex of the parabolic potential, (g) with the robot's propulsion vector directed tangentially in the counterclockwise direction, and (h) at the edge the robot's propulsion vector directed tangentially in the clockwise direction. [(i)-(l)] The same as in panels (e)-(h), but for $S$-robots.}
    \label{fig:Single_bot_PWM_20}
\end{figure*}

We start by considering the motion of a single self-propelled robot placed in a parabolic potential implemented with the help of a satellite dish, the same as in the swarm experiments discussed in Sec.~\ref{sec:Parabolic}. The dish has two different main axes $D_{1}=110$~cm and $D_{2}=120$~cm, and therefore its horizontal cross sections are ellipses. Such a soft localizing potential allows us to study the robot's motion for an arbitrary duration, with the robots neither escaping the area captured by the camera nor condensing at the boundary.

In our experiments, we vary the robot activity (defined by PWM), the center of inertia by setting it at points $O$ and $S$, and the initial robot placement configuration defined by the robot's position and angular orientation. PWM takes the values $10\%$ (the low activity) and $20\%$ (the high activity) corresponding to the robot motion velocities of approximately $14$ and $17$ for $O$-robots, and $7$~cm/s and $12$~cm/s for $S$-robots, respectively. Four initial robot placement configurations are considered: (i) the ``center'' configuration corresponding to the robot placed at the vertex of the parabolic potential and its propulsion vector oriented along the semiminor axis, Fig.~\ref{fig:Single_bot_PWM_20}(a); (ii) the ``edge'' configuration with the robot placed at the distance $20-35$~cm from the vertex at the semiminor axis and the propulsion vector directed towards the vertex point, Fig.~\ref{fig:Single_bot_PWM_20}(b); (iii) the ``orbit CCW'' configuration with the robot placed at the semi-minor axis at the distance $20-25$~cm from the vertex and the propulsion vector directed along the semimajor axis in such a manner that the robot moves in a CCW-direction, Fig.~\ref{fig:Single_bot_PWM_20}(c); and (iv) the ``orbit CW'' configuration when the robot is placed at the distance $20-30$~cm from the vertex at the semi-minor axis and the propulsion vector is directed along the semimajor axis resulting in a CW-directed robot's motion, Fig.~\ref{fig:Single_bot_PWM_20}(d). For each set of parameters, we repeated the experiment with three different robots whose motion has been recorded during $60$~s. To study the kinematic characteristics, we introduce the radius vector of robot $(\rho\cos\theta, \rho\sin\theta)$, defined in polar coordinates $\rho$ and $\theta$ with the origin at the vertex of the parabolic potential and the angle $\theta$ measured from the major semiaxis of the parabolic satellite dish. The radius vector points to the center of the robot's marker. We also introduce the robot orientation angle $\varphi$ between the robot propulsion vector and the radius vector.

Figures~\ref{fig:Single_bot_PWM_20}(e)-\ref{fig:Single_bot_PWM_20}(l) demonstrates the time evolution of the radius vector components $(\rho, \theta)$ and the orientation angle $\varphi$ of a single robot in the parabolic potential. The four previously described initial placement configurations of the robot moving with high activity ($\text{PWM}=20\%$) are studied for an $O$-robot and an $S$-robot with results displayed in Figs.~\ref{fig:Single_bot_PWM_20}(e)-\ref{fig:Single_bot_PWM_20}(h) and Figs.~\ref{fig:Single_bot_PWM_20}(i)-\ref{fig:Single_bot_PWM_20}(l), respectively. It is seen that there are two major types of individual robot motion, in accordance with Ref.~\cite{2019_Dauchot}: a car drifting-like \textit{climbing motion}, whose distinctive feature is the propulsion vector directed outwards the vertex of the parabolic potential and the average velocity directed tangentially to height isolines, and the \textit{orbiting motion} with the propulsion vector oriented tangentially to the elliptical trajectory of the robot.

The first type of motion is characterized by the orientation angle $\varphi$ close to $180^{\circ}$; see the dashed horizontal lines in Figs.~\ref{fig:Single_bot_PWM_20}(e)-\ref{fig:Single_bot_PWM_20}(l). Moreover, the radial coordinate $\rho$ and the polar angle $\theta$ change slowly (up to some fast fluctuations), with most of the curves corresponding to the robot completing a single revolution around the potential's vertex during the observed timespan [see the blue, green, and orange solid lines in Fig.~\ref{fig:Single_bot_PWM_20}(e) and the green solid line in Figs.~\ref{fig:Single_bot_PWM_20}(g), \ref{fig:Single_bot_PWM_20}(k), and \ref{fig:Single_bot_PWM_20}(l)] or even an incomplete rotation [the orange solid lines in Fig.~\ref{fig:Single_bot_PWM_20}(h) and Figs.~\ref{fig:Single_bot_PWM_20}(i)-\ref{fig:Single_bot_PWM_20}(l) and the green solid lines in Figs.~\ref{fig:Single_bot_PWM_20}(i) and \ref{fig:Single_bot_PWM_20}(j)]. The radial coordinate $\rho$ changes with the same oscillation period as $\theta$ corresponding to the robot moving at the same height along the elliptical trajectory characterized by different values of the semiminor and semimajor axes. Due to the ellipticity of the trajectory, oscillations with the same period can also be observed in the orientation angle $\varphi$, which would not exist if the trajectory was a perfect circle. The described climbing motion occurs for the $O$-robot, Figs.~\ref{fig:Single_bot_PWM_20}(e)-\ref{fig:Single_bot_PWM_20}(h) [the blue solid line in Fig.~\ref{fig:Single_bot_PWM_20}(e), the green solid lines in Figs.~\ref{fig:Single_bot_PWM_20}(e) and \ref{fig:Single_bot_PWM_20}(g), and the orange solid lines in Figs.~\ref{fig:Single_bot_PWM_20}(e) and ~\ref{fig:Single_bot_PWM_20}(h)] as well as for the $S$-robot [the green and orange solid lines in Figs.~\ref{fig:Single_bot_PWM_20}(i)-~\ref{fig:Single_bot_PWM_20}(l)]. The fast fluctuations mentioned previously, which are clearly manifested in the radial coordinate $\rho$ and the orientation angle $\varphi$, are related to the robot's oscillating up-and-down motion around a steady height level, see supplementary video~4. In addition, the rattling of $\varphi$ with frequencies greatly exceeding those of the oscillations of $\theta$ and an amplitude that may in some cases reach $25^{\circ}$ [see Fig.~\ref{fig:Single_bot_PWM_20}(e)] results from a noise in the robot's propulsion vector orientation due to the bristle-bot vibrating motion mechanism. It is seen that such rattling is only present in the orientation angle $\varphi$ but absent in the polar angle $\theta$.

The second type of single robot motion is characterized by fast oscillations of the radial coordinate $\rho$ accompanied by a sawtooth waveform in the polar angle $\theta$, both having the same period, Fig.~\ref{fig:Single_bot_PWM_20}. This signifies that the robot repeatedly traverses an elliptical trajectory, while the period of these oscillations is inversely proportional to the robot's tangential velocity. Moreover, oscillations with a more complicated waveform are observed in the orientation angle $\varphi$ that stays either in the vicinity of $90^\circ$ or $270^\circ$, indicating that the robot's propulsion vector direction is well aligned with its direction of motion along the elliptical trajectory. Thus, in contrast to the climbing motion described above, this second type of motion resembles the movement of a car running on an oval track by proper steering. Similarly to the previous motion type, this orbiting motion is observed for both the $O$-robots shown with the blue and orange solid lines in Figs.~\ref{fig:Single_bot_PWM_20}(f) and \ref{fig:Single_bot_PWM_20}(g) and the blue solid line in Fig.~\ref{fig:Single_bot_PWM_20}(h), and the $S$-robots shown with the blue solid lines in Figs.~\ref{fig:Single_bot_PWM_20}(i)-\ref{fig:Single_bot_PWM_20}(l).

In some cases, a hybrid type of a single robot motion can be observed, characterized by the orientation angle $\varphi$ taking the values between $180^{\circ}$ and $270^{\circ}$ and the robot making between two and three complete revolutions around the vertex per minute; see the solid green lines in Figs.~\ref{fig:Single_bot_PWM_20}(f)-\ref{fig:Single_bot_PWM_20}(h). The presence of such motion profiles with intermediate characteristics demonstrates that the two previously described regimes may be considered as limiting cases of a single type of motion corresponding to the robot moving periodically along an elliptical trajectory with a certain angle between the robot's velocity vector and its propulsion vector. Greater values of this angle correspond to more pronounced radial oscillations with a longer period.

To evaluate the role of collisions in the single-particle motion of robots, rotational excitation was applied in the form of instantaneous strikes to various points of the robot side surface far from its center of inertia. The strike causes a transient rotation, which damps in $\tau = 0.6\pm0.2$~s for $O$-robots and $\tau = 1.0\pm0.3$~s for $S$-robots. After that, the robots return to their normal movement pattern, that is, climbing or orbiting motion. Detailed analysis of these experiments is provided in the supplementary material~\cite{Supplement}. Such damping times are significantly smaller than other characteristic times in the system. Therefore, transport of angular momentum can be neglected due to its almost instantaneous damping by bristle-bots, and robot collisions that do not result in cluster formation are unlikely to play a significant role in micellization.

In summary, the motion profiles of $O$-robots and $S$-robots demonstrate similar properties. It is also seen from the $\rho(t)$ curves for the orbiting motion in Fig.~\ref{fig:Single_bot_PWM_20} that $O$-robots move along smaller ellipses compared to $S$-robots. Therefore, those of the $O$-robots that move slightly closer to the vertex of the parabolic potential than $S$-robots, are presumably more likely to collide with each other attempting to form clusters, some of which may end up being stable. However, as seen in Fig.~\ref{fig:Experiments_Tabl} and Table~\ref{tab:Potential_micelle_time}, micelles do not form in $O$-robot swarms. Thus, the increase in $O$-robots collisions seemingly leads to the formation of clusters that are either unstable, or stable but not contributing to (or even preventing) micelle formation. Therefore, the increased micellization of $S$-robots compared to $O$-robots cannot be explained by differences in the movement of a single particle, and collisions of at least two particles should be considered.

\subsection{Dominant stable two-particle cluster type defined by particle center of inertia location}
\label{sec:Two_Robots}

\begin{figure*}[tbp]
    \centering
    \includegraphics[width=\textwidth]{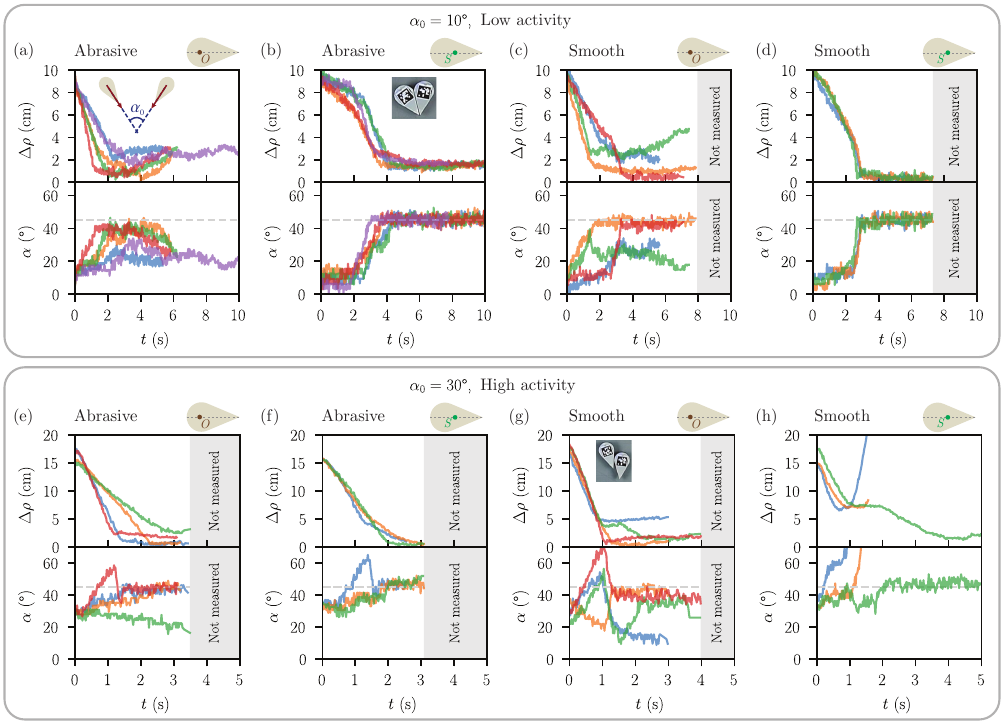}
    \caption{Dynamics of two-robot collisions for nonchiral self-propelled robots moving on a flat surface. [(a)-(d)] Time dependencies of the distance between the first and the second robots' noses $\Delta\rho$ and the angle between the robots' propulsion vectors $\alpha$. Robots move with the low vibration activity ($\text{PWM}=10\%$). Solid lines with different colors in each panel correspond to different realizations of the collision experiment with the same impact angle $\alpha_0=10^\circ$. Individual parameters for the panels are the following: (a) $O$-robots with abrasive lateral surfaces, (b) $S$-robots with abrasive lateral surfaces, (c) $O$-robots with smooth lateral surfaces, and (d) $S$-robots with smooth lateral surfaces. The gray-shaded areas in panels (c) and (d) denote the time in which the robots reach the region inaccessible for recording and turned off. The gray dashed line shows the angle $\alpha = 45^\circ$ which corresponds to the formation of a ``heart'' cluster; see Fig.~\ref{fig:OrderParam}(a) and the inset on panel (b). [(e)-(h)] The same as panels (a)-(d), but for the impact angle $\alpha_0=30^\circ$ and robots moving with the high vibration activity ($\text{PWM}=20\%$). The inset of Panel (g) shows the ``tandem'' cluster.}
    \label{fig:Two_Robots_PWM_20}
\end{figure*}

To address microscopic micellization mechanisms beyond the single-particle level, we proceed with considering the collisions of two self-propelled robots for different sets of parameters: low ($\text{PWM}=10\%$) and high ($\text{PWM}=20\%$) activity of the robots, abrasive and smooth lateral surfaces, and impact angles $\alpha_0 = \{10^\circ, 30^\circ, 60^\circ\}$; see Fig.~\ref{fig:Two_Robots_PWM_20}(a). Despite single-robot experiments were conducted in a shallow parabolic potential of a satellite dish, we consider nonchiral robots moving linearly on a plane when studying two-robot collisions, as controlling the impact angle for robots moving along elliptical or circular trajectories appears extremely challenging.

Figure~\ref{fig:Two_Robots_PWM_20} shows the key results for $O$- and $S$-robots with smooth and abrasive lateral surfaces moving with low and high activities and colliding at impact angles $\alpha_{0} = 10^\circ$ [Figs.~\ref{fig:Two_Robots_PWM_20}(a)-\ref{fig:Two_Robots_PWM_20}(d)] and $\alpha_{0} = 30^\circ$ [Figs.~\ref{fig:Two_Robots_PWM_20}(e)-\ref{fig:Two_Robots_PWM_20}(h)] while additional data for low activity, as well as for $\alpha_{0} = 60^\circ$ are considered in the supplementary material~\cite{Supplement}. For each set of parameters, we perform three to four experiments with different pairs of robots to ensure that the observed properties of scattering and cluster formation indeed represent the characteristic behavior of the considered system rather than the specific properties of certain robots. In the following, we focus on the evolution of the distance $\Delta \rho$ between the noses of the robots and the angle $\alpha$ between their propulsion vectors.

It is clearly observed that, regardless of the impact angle and the presence of an abrasive coating on their lateral surfaces, $S$-robots form stable (i.e., moving without any considerable shape variations) heart-shaped two-robot clusters after a collision,; see the inset in Fig.~\ref{fig:Two_Robots_PWM_20}(b) and supplementary video~5. As seen in Figs.~\ref{fig:Two_Robots_PWM_20}(b), \ref{fig:Two_Robots_PWM_20}(d), \ref{fig:Two_Robots_PWM_20}(f),and \ref{fig:Two_Robots_PWM_20}(h), formation of such clusters is indicated by an almost vanishing distance between the robots' noses and the angle between the robots' propulsion vectors tending to $45^\circ$ (the gray dashed lines in Fig.~\ref{fig:Two_Robots_PWM_20}). In particular, during the first stage (lasting approximately $1$~s) of the experiment, the angle $\alpha$ keeps its initial value $\alpha_{0}$, while the distance between robots' noses decreases linearly. After that, the second stage of a collision takes place, hallmarked by an abrupt change in the angle $\alpha$ from $\alpha_0$ to $\alpha_{\text{h}} = 45^{\circ}$ accompanied by the distance $\Delta \rho$ reaching $\Delta\rho_{\text{h}} = 0$~cm. Physically, this corresponds to the robots adjusting their propulsion vector directions by rotating around the point of contact until $\alpha_{\text{h}} = 45^{\circ}$ is reached. Finally, during the third stage, both $\alpha \approx 45^{\circ}$ and $\Delta\rho \approx 0$~cm remain nearly unchanged (up to some noise-like fluctuations), highlighting that two robots now propel forward in a continuous contact. The described ``heart'' clusters essentially represent quarter-micelles and can potentially contribute to the formation of half-micelles (consisting of four robots) after a collision between two such clusters.

The described scenario of ``heart'' cluster formation is most clearly observed in Figs.~\ref{fig:Two_Robots_PWM_20}(b) and \ref{fig:Two_Robots_PWM_20}(d) for the robots starting with $\alpha_{0} = 10^{\circ}$ which considerably differs from $\alpha_{\text{h}} = 45^{\circ}$, while for $\alpha_{0} = 30^{\circ}$ the transition region is much less pronounced, as seen in Figs.~\ref{fig:Two_Robots_PWM_20}(f) and \ref{fig:Two_Robots_PWM_20}(h). However, due to the increased technical difficulty in engineering the collisions between the robots for $\alpha_{0} = 30^{\circ}$ compared to $\alpha_{0} = 10^{\circ}$, there are considerable deviations between the target value of $\alpha_{0} = 30^{\circ}$ and the actual impact angle obtained. In particular, Fig.~\ref{fig:Two_Robots_PWM_20}(h) demonstrates $\alpha_{0}$ close to $35^{\circ}$ in all three experiments. Moreover, in some cases collisions of the robots do not result in cluster formation [see the blue and orange solid lines in Fig.~\ref{fig:Two_Robots_PWM_20}(h)].

In contrast, for $O$-robots the ``heart'' clusters are unstable, as seen in Figs.~\ref{fig:Two_Robots_PWM_20}(a) and \ref{fig:Two_Robots_PWM_20}(c). In particular, when $\alpha$ and $\Delta \rho$ reach their values of $\alpha_{\text{h}} = 45^{\circ}$ and $\Delta\rho_{\text{h}} = 0$~cm characteristic of the heart-shaped clusters, they fluctuate around their mean values for $0.5..1$~s. After that, $\Delta\rho$ starts growing until it reaches $\Delta\rho_{\text{t}} \approx 5$~cm, while $\alpha$ decreases to $\alpha_{\text{t}} \approx 15^{\circ}$, as seen for all experiments in Fig.~\ref{fig:Two_Robots_PWM_20}(a) and for the experiment shown with the green solid line in Fig.~\ref{fig:Two_Robots_PWM_20}(c). The meaning of these characteristic values of $\alpha$ and $\rho$ is uncovered in Fig.~\ref{fig:Two_Robots_PWM_20}(g). It turns out that $O$-robots support the formation of a different stable cluster type -- the ``tandem'' cluster -- in which one robot touches with its nose the round part of the other robot, as shown in the inset in Fig.~\ref{fig:Two_Robots_PWM_20}(g) and supplementary video~5. Such clusters can also travel with the values of $\alpha$ and $\Delta\rho$ retained, up to some fluctuations. Thus, the ``heart'' cluster in the case of $O$-robots is inherently metastable and eventually transforms into the asymmetric ``tandem'' cluster described above. Note that for higher values of the impact angle $\alpha_{0}$, the formation of such ``tandem'' clusters is more pronounced for robots with abrasive lateral surfaces, as seen from the comparison of Fig.~\ref{fig:Two_Robots_PWM_20}(e) and Fig.~\ref{fig:Two_Robots_PWM_20}(g).

To summarize, the asymmetric ``tandem'' clusters observed in collisions of $O$-robots do not represent quarter-micelles and are thus unlikely to contribute to micelle formation. In contrast, $S$-robots form symmetric ``heart'' clusters that can potentially contribute to formation of micelles. Thus, the observed emergence of micellization appears to be related (at least partially) to two-particle effects, in particular to the formation of stable quarter-micelle motile clusters.

\section{Discussion}
\label{sec:Discussion}


In this work, we experimentally demonstrate the emergence of micellization in self-propelled particle ensembles implemented by robotic swarms, which is governed by an interplay of particles' activity and their shape asymmetry, in contrast to micellization of surfactants in water solutions due to their inability to form hydrogen bonds~\cite{1962_Nemethy_PtI, 1962_Nemethy_PtII, 1962_Nemethy_PtIII}. To prevent the dominance of boundary condensation~\cite{2013_Giomi, 2018_Deblais} which would inevitably mask bulk phenomena such as micelle formation, we consider two scenarios: systems of chiral robots moving on a flat surface along circular orbits with diameters much larger than the size of an individual robot, yet considerably smaller than the diameter of the boundary, and nonchiral robots moving linearly but placed in a parabolic satellite dish creating a soft localizing potential.

As we demonstrate, the presence or absence of micellization is defined by the location of the robot center of inertia, which supports the numerical predictions~\cite{2021_Kruglov}. In particular, the robots can form micelles if their center of inertia is shifted towards the nose, while the micellization is suppressed for the center of inertia located close to the geometrical center. Such a behavior is clearly observed in both cases, for swarms of chiral robots on a flat surface and for swarms of nonchiral robots in a parabolic potential. The presence of such a localizing potential reduces the characteristic time of micelle formation and greatly enhances the mean lifetime of micelles.
However, such micellelike clusters in the considered system can break apart due to the interaction with single robots or other clusters, and are generally metastable, in contrast to micelles formed by surfactants. As a result, a transition to the fully micellar phase in the considered system appears unrealistic. Additional studies of the stability of individual micelles as well as their interaction with the rest of the robotic swarm are discussed in the supplementary material~\cite{Supplement}.

To underpin the microscopic physical mechanisms responsible for the formation of micelles on changing the location of the robot center of inertia, we performed two series of experiments addressing the motion characteristics of individual robots and the properties of two-robot collisions. Although the observed difference in the single-robot behavior cannot be directly linked to the emergence of micellization, our two-robot collision experiments clearly indicate the formation of two different types of stable two-robot clusters depending on the location of robot center of inertia. One of them represents a quarter of the micelle, thus being its constructive block. Therefore, the observed emergence of micellization is at least a two-particle effect.

The role of higher-order contributions and inertial effects~\cite{2018_Scholz, 2020_Leoni, 2019_Mandal} has not been considered in the present article, as arranging simultaneous three- or four-particle collisions in a laboratory setting and performing experiments to study collisions of individual robots with several-robot clusters appears to be a challenging task. It is a perspective direction for further study as it will allow one to uncover the role of many-particle microscopic effects in micelle formation and identify whether the two-particle processes considered in the present paper are the key factor. Moreover, various mechanisms might exist that can lead to a sufficient increase in the micelle stability that are mediated by the interaction of micelles and other clusters. For example, supplementary video~6 demonstrates an unusual herding of a complete micelle by a single robot roaming along the perimeter of the micelle that serves as a ``shepherd'' and continuously maintains the shape of the micelle, resulting in an anomalous lifetime of at least $238$~s (the cluster has formed at the timestamp $62$~s and remained stable throughout the experiment).

The second direction for further research is to consider swarms with nonzero net chirality in contrast to those studied in Sec.~\ref{sec:Chiral_Exp}. For chiral particles, the emergence of edge states has been reported in physical settings involving swarms of various constituents ranging from self-rotating~\cite{2020_Yang} or self-propelled~\cite{2020_Barois} robots to bacteria with small net chirality~\cite{2024_Li} or even dancing pairs of people~\cite{2024_Du}. In the context of robotic swarm micellization, the presence of such edge states could modify the rate of robot condensation at the boundary. Moreover, fully chiral micelles with a considerable angular momentum are expected to demonstrate stability properties and formation mechanisms different from those of nonchiral or slightly chiral micelles.

Development of an analytical theoretical description for the reported micellization is yet another direction for further research. The system considered in the present paper is inherently nonequilibrium as the robots constantly convert the energy stored in their batteries to kinetic energy. It is also not clear whether an effective temperature can be properly defined in such a system~\cite{2008_Loi,2015_Takatori,2022_Sanjay,2022_Boudet,2024_Caprini}. This renders the approach of considering micellization as a reversible chemical reaction at equilibrium~\cite{1955_Phillips,2014_Nagarajan} (which is conventionally used to derive the free energy of micellization) inapplicable to our system. Moreover, in surfactant solutions, the micelle concentration increases as more surfactant is added to the mixture, in accordance with Le Chatelier's principle. However, for the reported micellization, increasing the robot concentration in some cases instead leads to a decrease in the number of micellelike clusters observed for abrasive $S$-robots, as is discussed in detail in the supplementary material~\cite{Supplement}. Moreover, a further increase in the number of robots is expected to completely prevent micellization due to the jamming transition~\cite{2019_Barois}.

Another prospective direction is the implementation of the reported micellization at smaller scales, for example, with millimeter-size particles propelled by Marangoni flows~\cite{2021_Shabaniverki,2022_Kumar}, and at the microscale, where new interaction mechanisms come into play, such as increased van der Waals forces. In terms of microscale implementations, Janus particles appear to be the most promising, as they allow one to change the activity of particles by an external illumination~\cite{2019_Villa, 2013_Palacci, 2016_Dai}. Moreover, self-propelled Janus particles with various asymmetric shapes have been synthesized~\cite{2016_Dai, 2020_Shah, 2022_Liu}, making the proposed design feasible.

Moreover, there is a wide range of possibilities in addressing complex external potentials~\cite{2021_Wang} and robot motion profiles~\cite{2022_Zhang} in order to develop novel control paradigms for robotic swarms~\cite{2019_Li}, which can involve tools such as reinforcement learning~\cite{2023_Zion}. Finally, the reported micellization, once reproduced at the microscale, might find practical applications in the areas of microfluidics, microstructuring of materials, and waste removal. For example, micropumps based on self-rotating snowman-shaped particles have been demonstrated~\cite{2015_vandenBeld}. The use of self-propelled microparticles has also been proposed to synthesize self-organized random lasers~\cite{2022_Trivedi}, microplastics removal~\cite{2022_Li, 2022_Won}, and enhanced catalytic decomposition of medical masks~\cite{2023_Kim}.

\begin{acknowledgments}

We acknowledge fruitful discussions with Dr. Timofey Kruglov, who brought the idea of active - matter micellization to our attention; Dr. Alexander Borisov, Prof. Mikhail Lapine; and Prof. Anton Souslov. The work is supported by the Russian Science Foundation (project 24-79-10314).

\end{acknowledgments}


%

\end{document}


\title{Supplemental Material\\~\\Micellization in active matter of asymmetric self-propelled particles: Experiments}

\author{Anastasia~A.~Molodtsova}
\email{a.molodtsova@metalab.ifmo.ru}
\affiliation{School of Physics and Engineering, ITMO University, 197101 Saint Petersburg, Russia}

\author{Mikhail~K.~Buzakov}
\affiliation{School of Physics and Engineering, ITMO University, 197101 Saint Petersburg, Russia}

\author{Oleg~I.~Burmistrov}
\affiliation{School of Physics and Engineering, ITMO University, 197101 Saint Petersburg, Russia}

\author{Alina~D.~Rozenblit}
\affiliation{School of Physics and Engineering, ITMO University, 197101 Saint Petersburg, Russia}

\author{Vyacheslav~A.~Smirnov}
\affiliation{School of Physics and Engineering, ITMO University, 197101 Saint Petersburg, Russia}

\author{Daria~V.~Sennikova}
\affiliation{School of Physics and Engineering, ITMO University, 197101 Saint Petersburg, Russia}

\author{Vadim~A.~Porvatov}
\affiliation{University of Amsterdam, 1098 XH Amsterdam, Netherlands}
\affiliation{National University of Science and Technology ``MISIS'', Moscow 119991, Russia}

\author{Ekaterina~M.~Puhtina}
\affiliation{School of Physics and Engineering, ITMO University, 197101 Saint Petersburg, Russia}

\author{Alexey~A.~Dmitriev}
\affiliation{School of Physics and Engineering, ITMO University, 197101 Saint Petersburg, Russia}

\author{Nikita~A.~Olekhno}
\email{nikita.olekhno@metalab.ifmo.ru}
\affiliation{School of Physics and Engineering, ITMO University, 197101 Saint Petersburg, Russia}

\date{\today}

\maketitle

\tableofcontents


\section{Experimental setup}
\label{sec:Experimental_Setup}

\begin{figure*}[tbp]
    \centering
    \includegraphics[width=145mm]{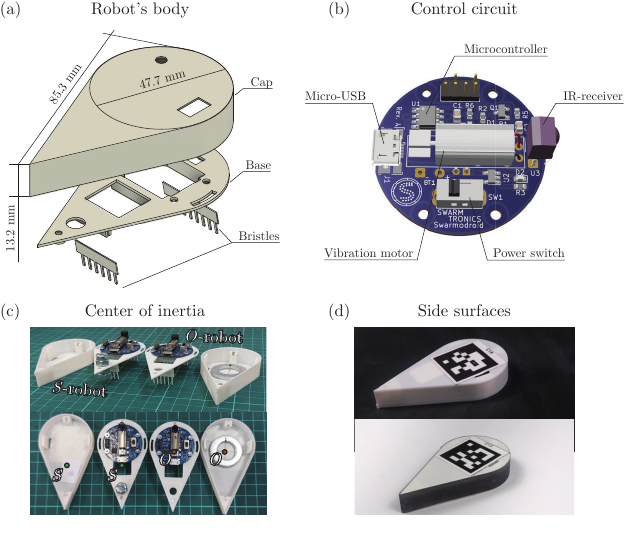}
    \caption{(a) Robot's body design. (b) Electric circuit as a PCB featuring a microcontroller, an IR receiver, a power switch, a vibration motor, and a micro-USB for charging. The battery is placed on the bottom side of the board. (c) Photographs demonstrating the $S$-robot with a center of inertia shifted towards the nose of the robot with the help of an M6 screw and three M6 nuts (left), as well as the $O$-robot with the center of inertia located near the center of the circular part having the same mass as the $S$-robot due to an M18 washer placed at the $O$ point (right). (d) Caps with smooth a side surface (top) and side surfaces covered with abrasive (bottom).}
    \label{figS:Experimental_setup}
\end{figure*}

Each robot consists of a PLA-plastic body and a control electrical circuit. Body construction includes a cap, a base and a couple of bristles, Fig.~\ref{figS:Experimental_setup}(a). All of these details were implemented via FDM (fused deposition modeling) 3D printing technology. The diameter of the circle that forms the teardrop shape of the cap is equal to $47.7$~mm, the length of the entire piece is $85.3$~mm, and the height is $13.2$~mm. The base of thickness $1.5$~mm incorporates a section for the battery and four circular holes to fasten the control circuit with M2 screws and nuts. The base is also provided with a circular hole in the anterior part of the base for a steel DIN 84 M6$\times$16 screw and three DIN 934 M6 steel nuts with a total mass of $10$~g to change the location of the center of inertia [Fig.~\ref{figS:Experimental_setup}(c)]. The cap and the base are fastened together by pairs of cylinder-shaped neodymium magnets of sizes $3 \times 2$~mm placed inside grooves of the corresponding size in the front and back parts of each piece. The robot is standing on two legs with a height of $10$~mm, each of which is a planar brush, i.e., an array of bristles with a rectangular shape with sizes $5.0 \times 0.8$~mm connected by a 5-mm-wide solid belt, 3D printed with a single layer PLA in one run. The flexibility of the bristles is facilitated by the thinness of the brushes, which is $0.4$~mm. The brushes are clamped in narrow grooves at the bottom of the base, inclined at an angle $80^{\circ}$ to the base plane. The front brush consists of $6$ bristles, while the rear brush has $8$ bristles. To emulate different values of friction coefficient between robots, we cover the side surface of the robots' caps with P800 sandpaper to achieve high friction, or leave the caps uncovered to facilitate low friction, as depicted in Fig.~\ref{figS:Experimental_setup}(d). To increase the mass of $O$-robots and make it the same as for $S$-robots while maintaining the center of inertia at point $O$, a DIN125A M18 steel washer with a mass of $10$~g is glued to the robot cap at point $O$ using cyanoacrylate adhesive.

The movement of the robots is captured using a Sony ZV-E10 camera set to a 50~fps frame rate. Each robot is provided with an individual ArUco marker to track its position using the OpenCV library, allowing to extract all kinematic characteristics of each robot during the video post-processing.

The electric circuit responsible for robot propulsion is implemented as a printed circuit board (PCB) carrying an ATTiny13A microcontroller, a QX-6A-1 vibration motor, a Vishay TSOP4838 infrared (IR) receiver, a Robiton LP601120 lithium-ion battery, an SS12D07 power switch, a micro-USB for charging, as well as passive auxiliary components such as surface-mounted capacitors and resistors, Fig.~\ref{figS:Experimental_setup}(b). Robots can be given commands by an IR remote control device via the NEC protocol, which also offers external control over the vibration intensity of their motors by setting the duty cycle of the pulse width modulation of the motor voltage to $0\%, 5\%, 10\%, \ldots, 50\%$, according to the command received from the remote control device. Using a remote control to set the vibration intensity also ensures that all robots start moving from their initial positions at the same moment. We note that the power switch is only provided to eliminate unwanted battery discharge during long-term storage of the robots and always remains switched on while a series of experiments are performed.

\section{The choice of abrasive materials for side surfaces of robots}
\label{sec:Friction}

\begin{figure*}[tbp]
    \centering
    \includegraphics[width=170mm]{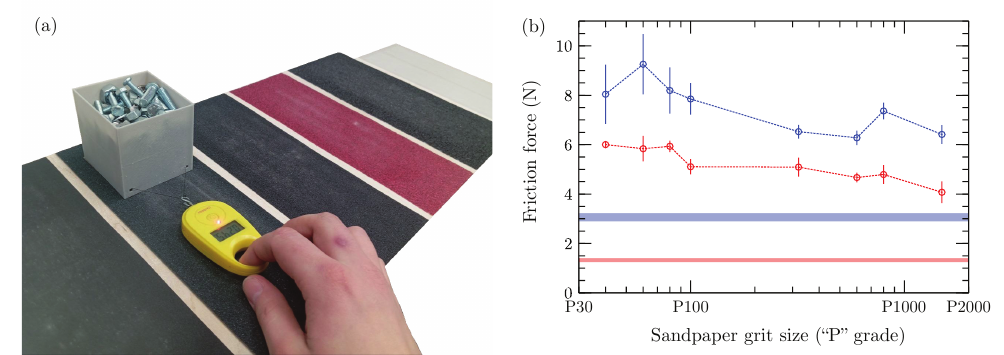}
    \caption{(a) The experimental setup for sliding and static friction forces measurements. (b) The dependence of the static (the blue markers line) and dynamic (the red markers) friction forces between the sandpaper-covered flat surface and the sandpaper-covered box on the sandpaper grit according to the ``P'' grade. The markers and the error bars show the mean values and the dispersion over $10$ measurements, dashed lines serve as a guide for an eye, and the horizontal solid lines correspond to the mean values of the static (the red line) and the sliding (the blue line) friction forces between a flat surface and a box both covered by sheets of PLA. The thickness of horizontal lines denotes the dispersion over $10$ measurements.}
    \label{figS:Friction}
\end{figure*}

To assess how friction between the side surfaces of the robots affects the formation of micellized clusters, we start by characterizing several commonly accessible abrasive materials to select the most suitable ones. In particular, we consider textile-based abrasive papers with a range of grit sizes: P40, P60, P80, P100, P320, P600, P800, and P1500. To evaluate the friction coefficient for the contact of two surfaces covered by a given abrasive, we measure the friction force between a flat surface of a table and a moving box filled with fasteners with a total mass of $1000$~g, both covered by the respective material, using a digital spring scale graduated in kilogram-forces (four significant digits, maximum allowed force $10$~kgf), as shown in Figure~\ref{figS:Friction}(a). For each type of abrasive paper, we perform a series of $10$ experiments and evaluate the mean values and dispersions. The results for the measured static friction force that is determined by the minimal value required to initiate box movement and the dynamic friction force that indicates the force necessary to maintain box movement without acceleration are presented for different magnitudes of the sandpaper grit size in Figure~\ref{figS:Friction}(b). It is seen that both the static and dynamic friction forces do not change significantly for different sandpaper grit sizes. However, larger grit sizes (especially P40 and P60) correspond to larger scales of geometrical features at the sandpaper surface, which, in turn, change the effective size of the robots and the flatness of their side surfaces. Thus, for our experiments, we selected the abrasive paper P800. The horizontal lines in Figure~\ref{figS:Friction}(b) show the friction coefficient of PLA, which corresponds to robot caps without an abrasive cover. Compared to P800 sandpaper, the friction coefficient of PLA is approximately $2.5$~times lower. Thus, covering the robots with an abrasive paper indeed increases the friction between the robots significantly.

\begin{figure*}[tbp]
    \centering
    \includegraphics[width=170mm]{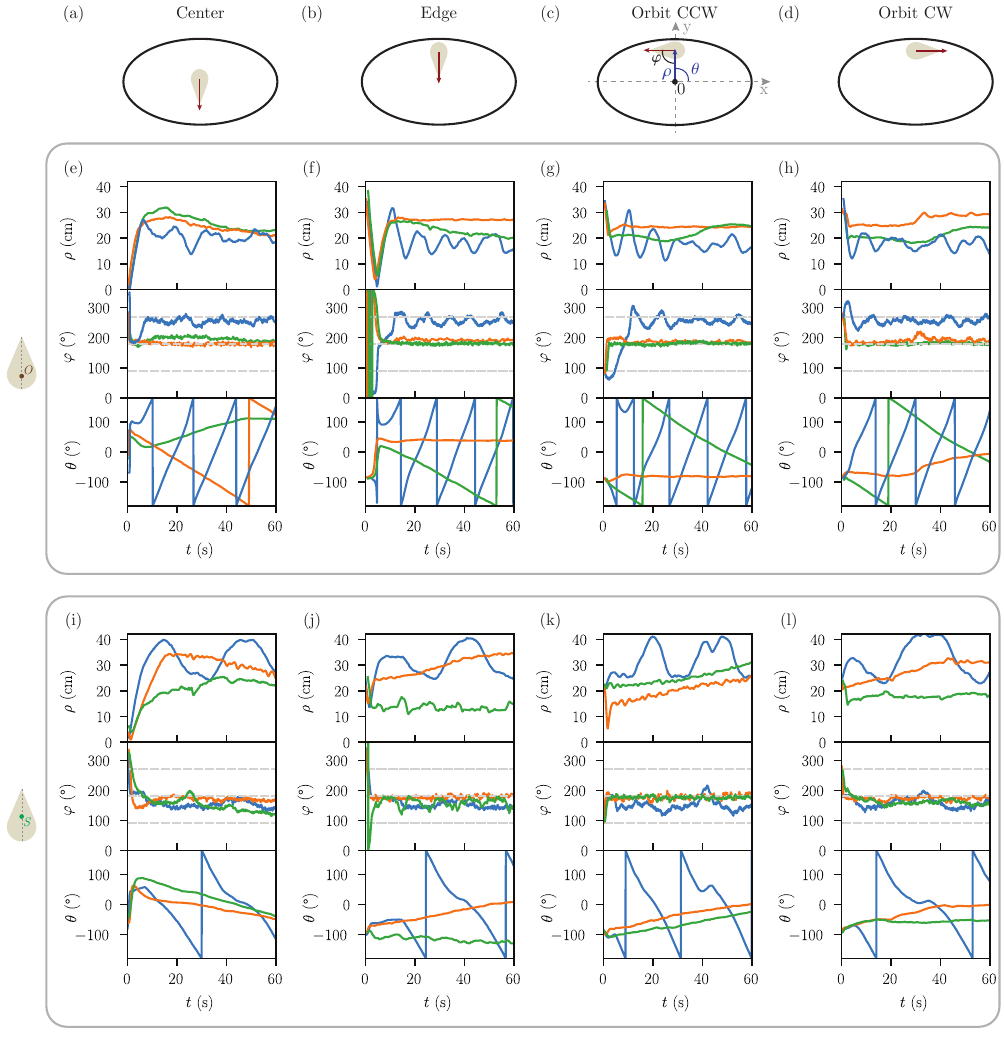}
    \caption{Single robot motion in a parabolic potential. (a)-(d) Schematic representation of the initial positions and orientations (gray outline) along with the propulsion directions (red arrows) of single robots in the parabolic satellite dish (black outline). Panel (c) also shows the polar coordinates $\rho$ and $\theta$ used to describe the radius vector, as well as the robot orientation angle $\varphi$. (e) Time dependencies of the radius vector components ($\rho$, $\theta$) and the orientation angle $\varphi$ of an $O$-robot in a parabolic potential. The blue, green, and orange solid lines correspond to three different robots, each moving with high activity ($\text{PWM}=10\%$), starting from the vertex of parabolic potential, as shown in Panel (a). The gray dashed lines in the middle panel indicate the angles $90^\circ$, $180^\circ$, and $270^\circ$. (f)-(h) The same as Panel (e), but for the initial positions and orientations of the robots at the edge of the parabolic potential corresponding to the ones shown in Panels (b)-(d): (f) with the robot's propulsion vector directed to the vertex of the parabolic potential, (g) with the robot's propulsion vector directed tangentially in the counter-clockwise direction, and (h) at the edge the robot's propulsion vector directed tangentially in the clockwise direction. (i)-(l) The same as in Panels (e)-(h), but for $S$-robots.}
    \label{figS:single_bot_PWM_10}
\end{figure*}

\begin{figure*}[tbp]
    \centering
    \includegraphics[width=170mm]{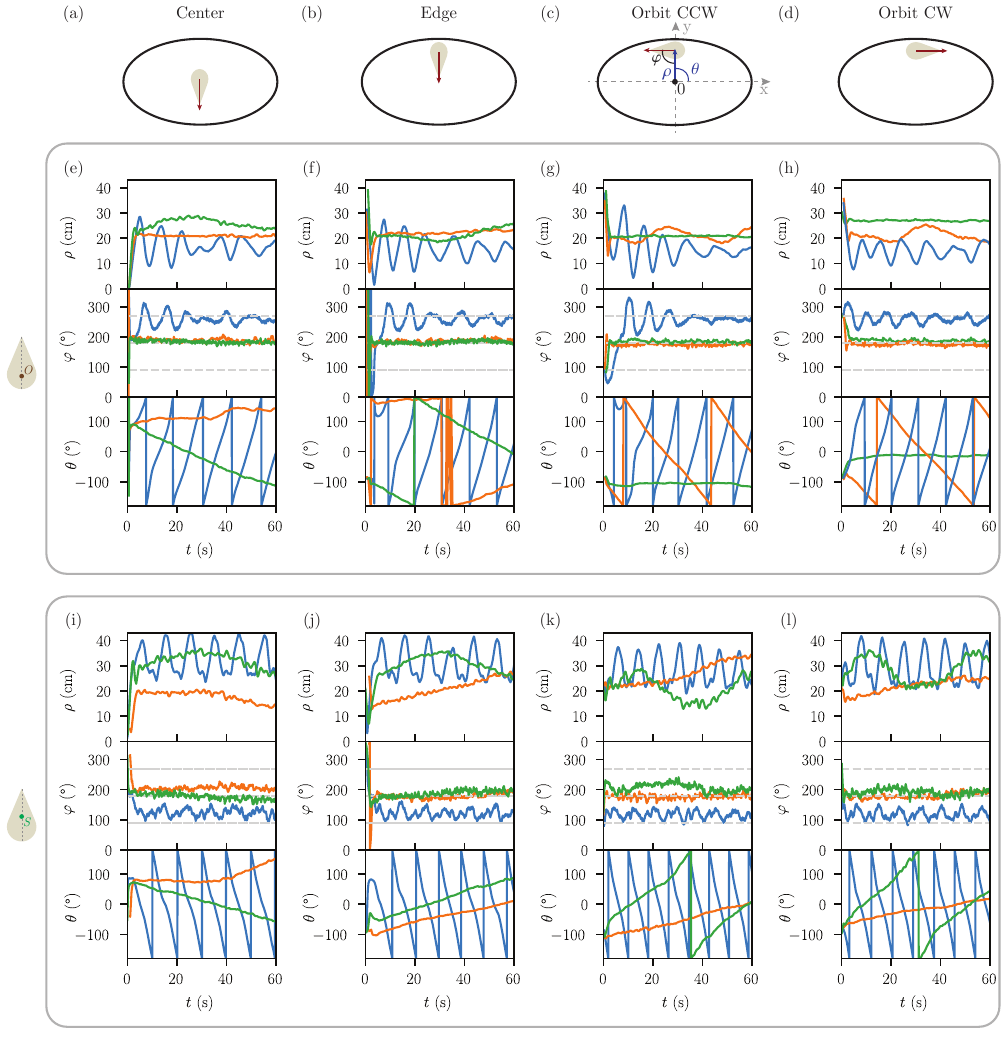}
    \caption{Single robot motion in a parabolic potential. (a)-(d) Schematic representation of the initial positions and orientations (gray outline) along with the propulsion directions (red arrows) of single robots in the parabolic satellite dish (black outline). Panel (c) also shows the polar coordinates $\rho$ and $\theta$ used to describe the radius vector, as well as the robot orientation angle $\varphi$. (e) Time dependencies of the radius vector components ($\rho$, $\theta$) and the orientation angle $\varphi$ of an $O$-robot in a parabolic potential. The blue, green, and orange solid lines correspond to three different robots, each moving with high activity ($\text{PWM}=20\%$), starting from the vertex of parabolic potential, as shown in Panel (a). The gray dashed lines in the middle panel indicate the angles $90^\circ$, $180^\circ$, and $270^\circ$. (f)-(h) The same as Panel (e), but for the initial positions and orientations of the robots at the edge of the parabolic potential corresponding to the ones shown in Panels (b)-(d): (f) with the robot's propulsion vector directed to the vertex of the parabolic potential, (g) with the robot's propulsion vector directed tangentially in the counter-clockwise direction, and (h) at the edge the robot's propulsion vector directed tangentially in the clockwise direction. (i)-(l) The same as in Panels (e)-(h), but for $S$-robots.}
    \label{figS:single_bot_PWM_20}
\end{figure*}

\begin{figure*}[tbp]
    \centering
    \includegraphics[width=170mm]{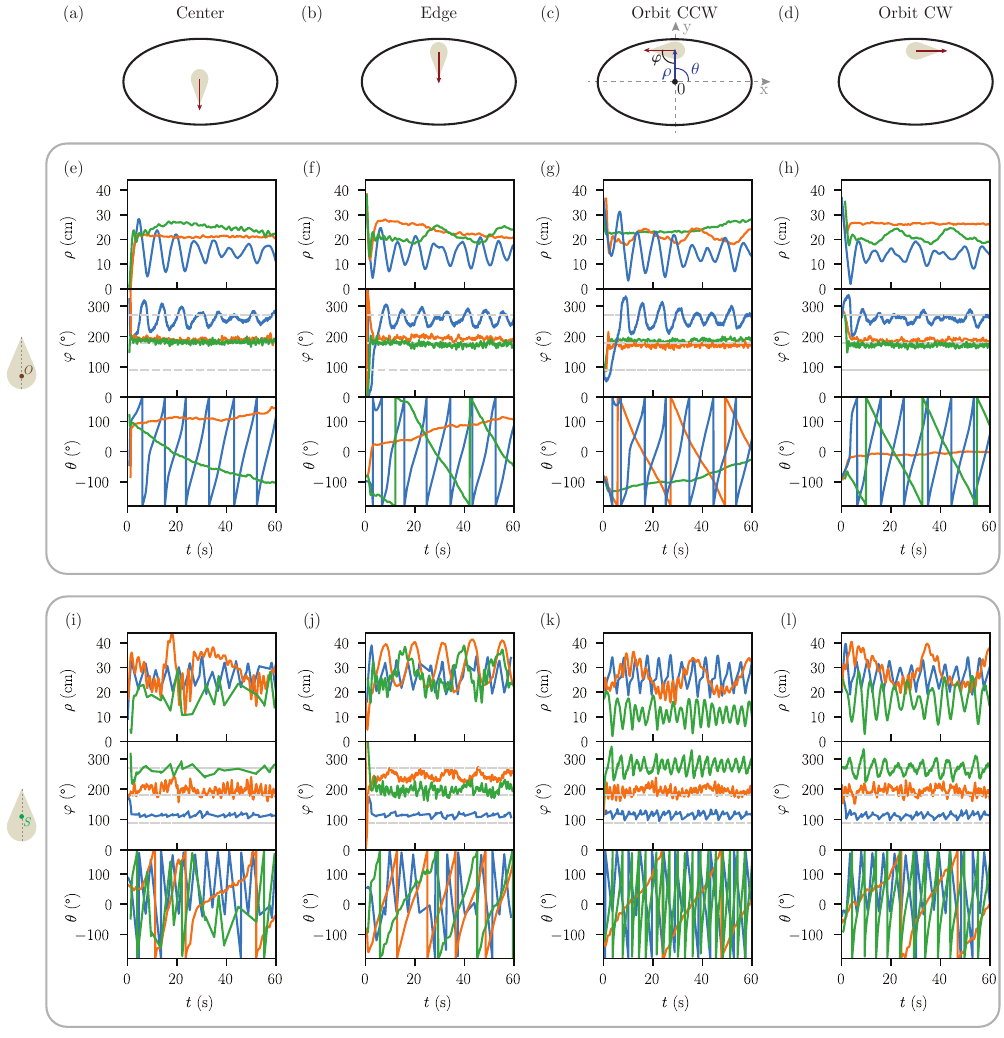}
    \caption{Single robot motion in a parabolic potential. (a)-(d) Schematic representation of the initial positions and orientations (gray outline) along with the propulsion directions (red arrows) of single robots in the parabolic satellite dish (black outline). Panel (c) also shows the polar coordinates $\rho$ and $\theta$ used to describe the radius vector, as well as the robot orientation angle $\varphi$. (e) Time dependencies of the radius vector components ($\rho$, $\theta$) and the orientation angle $\varphi$ of an $O$-robot in a parabolic potential. The blue, green, and orange solid lines correspond to three different robots, each moving with high activity ($\text{PWM}=30\%$), starting from the vertex of parabolic potential, as shown in Panel (a). The gray dashed lines in the middle panel indicate the angles $90^\circ$, $180^\circ$, and $270^\circ$. (f)-(h) The same as Panel (e), but for the initial positions and orientations of the robots at the edge of the parabolic potential corresponding to the ones shown in Panels (b)-(d): (f) with the robot's propulsion vector directed to the vertex of the parabolic potential, (g) with the robot's propulsion vector directed tangentially in the counter-clockwise direction, and (h) at the edge the robot's propulsion vector directed tangentially in the clockwise direction. (i)-(l) The same as in Panels (e)-(h), but for $S$-robots.}
    \label{figS:single_bot_PWM_30}
\end{figure*}

\section{Single-robot motion in different regimes}
\label{sec:Single_robots_Supp}

We studied the dynamics of a single robot in a parabolic potential by varying the following parameters: robot activity (PWM), center of inertia location, and the robot's initial placement and orientation, as shown in Figs.~\ref{figS:single_bot_PWM_10}(a)-(d). As discussed in the work~\cite{2019_Dauchot}, there are two main types of individual robot motion: climbing motion and orbiting motion.

Figure~\ref{figS:single_bot_PWM_10} shows the evolution of the polar angle $\theta$, the displacement from the center of potential $\rho$, and the orientation angle $\varphi$ for a single robot moving at PWM$=10\%$. The solid orange and green lines in Figs.~\ref{figS:single_bot_PWM_10}(e),(f),(h),(l), the orange solid line in Fig.~\ref{figS:single_bot_PWM_10}(g), and the blue solid line in Fig.~\ref{figS:single_bot_PWM_10}(h) correspond to the climbing motion. In turn, the solid blue lines in Fig.~\ref{figS:single_bot_PWM_10}(f) and the solid blue and green lines in Fig.~\ref{figS:single_bot_PWM_10}(g) correspond to the orbiting motion. Moreover, there is an intermediate regime corresponding to the blue solid lines in Figs.~\ref{figS:single_bot_PWM_10}(e),(i)-(l).

Next, an analogous study is presented in Figure~\ref{figS:single_bot_PWM_20} for robots moving at a higher activity PWM$=20\%$ with the same set of initial positions and orientations, Fig.~\ref{figS:single_bot_PWM_20}(a)-(d). The blue, green, and orange solid lines in Fig.~\ref{figS:single_bot_PWM_20}(e), the green solid lines in Figs.~\ref{figS:single_bot_PWM_20}(i),(j), and the orange solid line in Figs.~\ref{figS:single_bot_PWM_20}(h),(i)-(l) correspond to the climbing motion. Meanwhile, the blue and orange solid lines in Figs.~\ref{figS:single_bot_PWM_20}(f),(g), the blue solid lines in Fig.~\ref{figS:single_bot_PWM_20}(h), and the blue solid lines in Figs.~\ref{figS:single_bot_PWM_20}(i)-(l) correspond to the orbiting motion. Moreover, an intermediate regime is also observed, as demonstrated by the green solid lines in Figs.~\ref{figS:single_bot_PWM_20}(f)-(h),(k),(l).

Finally, the case of a high activity PWM$=30\%$ is addressed in Figure~\ref{figS:single_bot_PWM_30}. As shown in Fig.~\ref{figS:single_bot_PWM_30}(a)-(d), the set of initial robot configurations is the same. The orange solid lines in Fig.~\ref{figS:single_bot_PWM_30}(i),(k)-(l), and the green solid line in Fig.~\ref{figS:single_bot_PWM_30}(e),(j) correspond to the climbing motion. In turn, the blue solid lines in Fig.~\ref{figS:single_bot_PWM_10}(e)-(g),(i)-(l) and the green solid lines in Fig.~\ref{figS:single_bot_PWM_30}(i),(k)-(l) correspond to the orbiting motion. The intermediate regime is demonstrated by the orange solid lines in Fig.~\ref{figS:single_bot_PWM_30}(e)-(h),(j), and the green and orange solid lines in Fig.~\ref{figS:single_bot_PWM_30}(f)-(h),(j).


\section{Single-robot motion under rotational excitation}
\label{sec:Rot_Supp}

To explore how collisions with other robots that do not result in cluster formation change the trajectory of single $O$- and $S$-robots and estimate the role of angular momentum transfer, we perform two series of experiments with additional rotational excitation of a robot that emulates such a collision. The rotational excitation was applied by inducing a torque on the robot instantaneously using a cue stick to strike it at a point shifted away from the center of inertia. The impact strength was chosen so as to make the robot rotate visibly, although collisions between robots occurring in the experiments considered in the main text did not induce torques large enough to make the robots rotate.

In the first set of experiments, we study the motion of individual $O$- and $S$-robots on a flat surface. In such a scenario, the robots quickly reach the boundary of a test area, similarly to the experiments on two-robot collisions. Thus, their motion is analyzed only by considering the robot orientation angle $\varphi$ and is recorded for a relatively short period of several seconds. 

\begin{figure*}[tbp]
    \centering
    \includegraphics{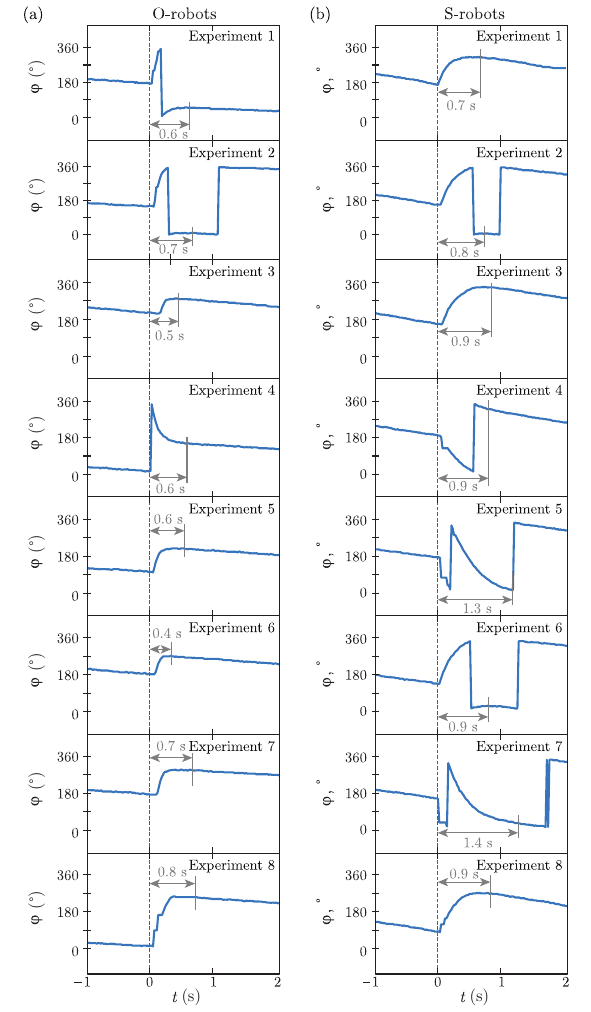}
    \caption{Movement of a single robot subject to instantaneous rotational excitation on a flat surface. Time dependencies of the orientation angle $\varphi$ of an $O$-robot (a) or an $S$-robot (b) moving on a flat surface and subject to instantaneous rotational excitation. The robot is moving with high activity ($\text{PWM}=20\%$), starting from the edge of parabolic potential. The red dashed lines denote the moments when a rotational excitation is applied.}
    \label{figS:RotationExcitationFlatSurface}
\end{figure*}

The results of these experiment are shown in Fig.~\ref{figS:RotationExcitationFlatSurface} for both $O$- and $S$-robots in Figs.~\ref{figS:RotationExcitationFlatSurface}(a)-(c) and Figs.~\ref{figS:RotationExcitationFlatSurface}(d)-(f) respectively. They demonstrate a similar behavior: the impact causes a transient rotation that damps in $\tau = 0.6\pm0.2$~s for $O$-robots and $\tau = 1.0\pm0.3$~s for $S$-robots, after which the robots return to their normal movement. This behavior does not depend on the force and point of impact. Such damping times are significantly less than other characteristic times in our system. Therefore, transport of angular momentum can be neglected due to its almost instantaneous damping by bristle-bots, and robot collisions that do not result in cluster formation are unlikely to play a significant role in micellization.

\begin{figure*}[tbp]
    \centering
    \includegraphics[width=170mm]{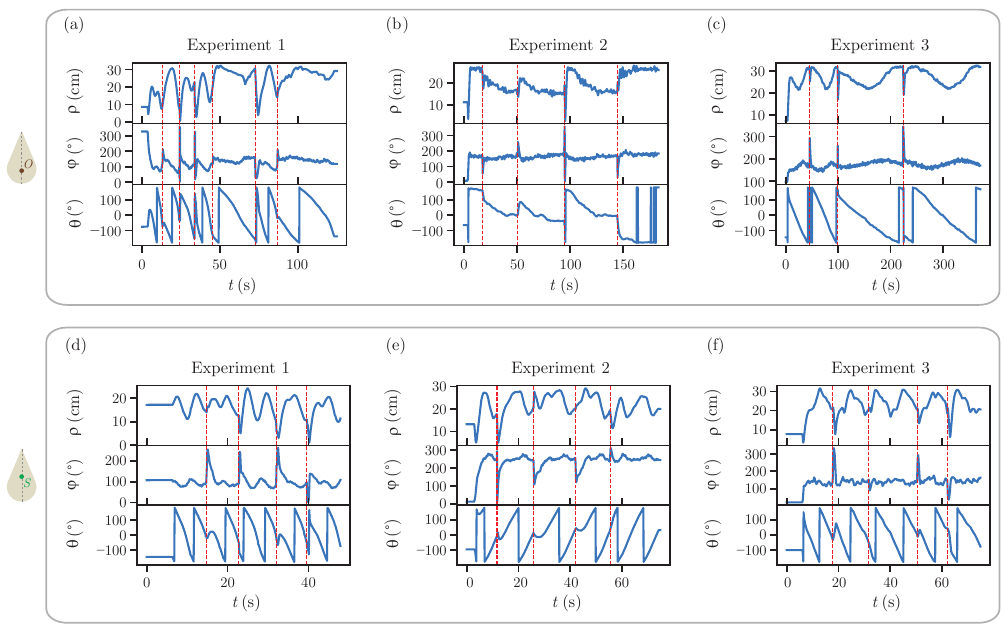}
    \caption{Motion of a single robot subject to an instantaneous rotational excitation inside a parabolic potential. (a)-(c) Time dependencies of the radius vector components ($\rho$, $\theta$) and the orientation angle $\varphi$ of an $O$-robot in a parabolic potential subject to instantaneous rotational excitation. The robot is moving with high activity ($\text{PWM}=20\%$), starting from the edge of parabolic potential. The red dashed lines indicate the moments when the rotational excitation is applied. (d)-(f) The same as Panels (a)-(c), but for an $S$-robot.}
    \label{figS:RotationExcitationParabolicDish}
\end{figure*}

Next, we consider the motion of individual $O$- and $S$-robots in a parabolic potential likewise to Figs.~\ref{figS:single_bot_PWM_10}--\ref{figS:single_bot_PWM_30}, but with instantaneous rotational excitation applied several times during an experiment in the same fashion as in Fig.~\ref{figS:RotationExcitationFlatSurface}. The results are demonstrated in Fig.~\ref{figS:RotationExcitationParabolicDish}. Similarly to Fig.~\ref{figS:RotationExcitationFlatSurface}, rotational excitation only results in a short period of transient rotation, after which the robot returns to one of the motion patterns demonstrated in Figs.~\ref{figS:single_bot_PWM_10}--\ref{figS:single_bot_PWM_30}, that is either climbing or orbiting motion.

Thus, it is seen that in a parabolic potential as well as on a flat surface, both $O$-robots and $S$-robots dissipate any transferred angular momentum almost instantly (compared to characteristic times of other processes in the system). On a flat surface, a collision simply leads to a robot changing its motion direction without obtaining any rotational component. In a parabolic potential, even the direction of robot motion appears robust to such an excitation, as seen in all discussed experiments in Fig.~\ref{figS:RotationExcitationParabolicDish}.

\section{Two-robot collisions with different parameters}
\label{sec:Two_robots_Supp}

As outlined in the main text, we consider the collision of two robots for different parameters. In the experiments, we vary the initial angle between the propulsion vectors of the robots $\alpha_0$, the robot activity (PWM), the location of the center of inertia, and the friction coefficient between the robots by coating their side surfaces with abrasives.

\begin{figure*}[tbp]
    \centering
    \includegraphics[width=170mm]{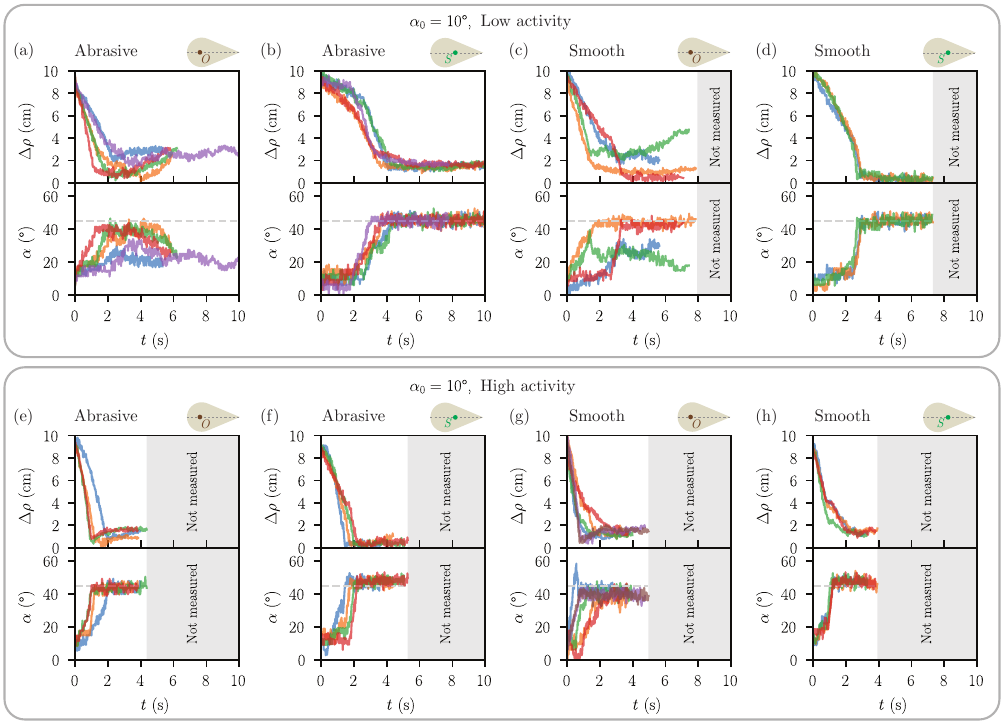}
    \caption{Kinematics of two-robot collisions for non-chiral self-propelled robots moving on a flat surface with the initial angle between the robots' propulsion vectors $\alpha_0=10^\circ$. (a)-(d) Time dependencies of the distance between the first and the second robots' noses $\Delta\rho$ and the angle between the robots' propulsion vectors $\alpha$. Solid lines with different colors in each panel correspond to different realizations of the collision experiment with the same initial angle between the robots moving at $\text{PWM}=10\%$. Individual parameters for the panels are the following: (a) $O$-robots with abrasive side surfaces, (b) $S$-robots with abrasive side surfaces, (c) $O$-robots with smooth side surfaces, and (d) $S$-robots with smooth side surfaces. The shaded gray areas in Panels (a), (c), and (d) denote correspond to robots reaching the region inaccessible for recording. The gray dashed line shows the angle $\alpha = 45^\circ$ characteristic of a ``heart'' cluster. (e)-(h) The same as Panels (a)-(d), but for the robots moving at $\text{PWM}=20\%$.}
    \label{figS:collision_angle_10}
\end{figure*}

Figure~\ref{figS:collision_angle_10} shows the evolution of the distance between the noses of the first and second robots $\Delta \rho$ and the angle between the robot propulsion vectors $\alpha$ with the initial angle $\alpha_0 = 10^\circ$. Two cases are considered: low robot activity with PWM$=10\%$ [Figs.~\ref{figS:collision_angle_10}(a)-(d)] and a high robot activity with PWM$=20\%$ [Figs.~\ref{figS:collision_angle_10}(e)-(h)]. It is observed that for all realizations with $S$-robots, Figs.~\ref{figS:collision_angle_10}(b),(d),(f),(h), the stable heart-shaped clusters are formed, as indicated by the almost zero distance between the first and second robot noses $\Delta\rho$ and the angle between the robot propulsion vectors $\alpha$ reaching the plateau of $45^\circ$. For the realizations with $O$-robots, the formation of such a cluster is not observed, Figs.~\ref{figS:collision_angle_10}(a),(c),(e),(g). However, the blue and orange solid lines in Fig.~\ref{figS:collision_angle_10}(a) almost reach the corresponding plateaus, but the angle $\alpha$ does not approach the required value of $45^\circ$. In several experiments, robots do not form a stable cluster upon a collision. For example, the solid green line in Figs.~\ref{figS:collision_angle_10}(c) corresponds to the robots touching at $t \approx 2$~s, but then again increasing their relative distance, i.e., the corresponding experiments represent a scattering of two robots. The solid green and orange lines in Figs.~\ref{figS:collision_angle_10}(g) represent the formation of unstable ``heart'' (symmetric) and ``tandem'' (asymmetric) two-robot clusters, respectively, which begin to break apart after their formation.

\begin{figure*}[tbp]
    \centering
    \includegraphics[width=170mm]{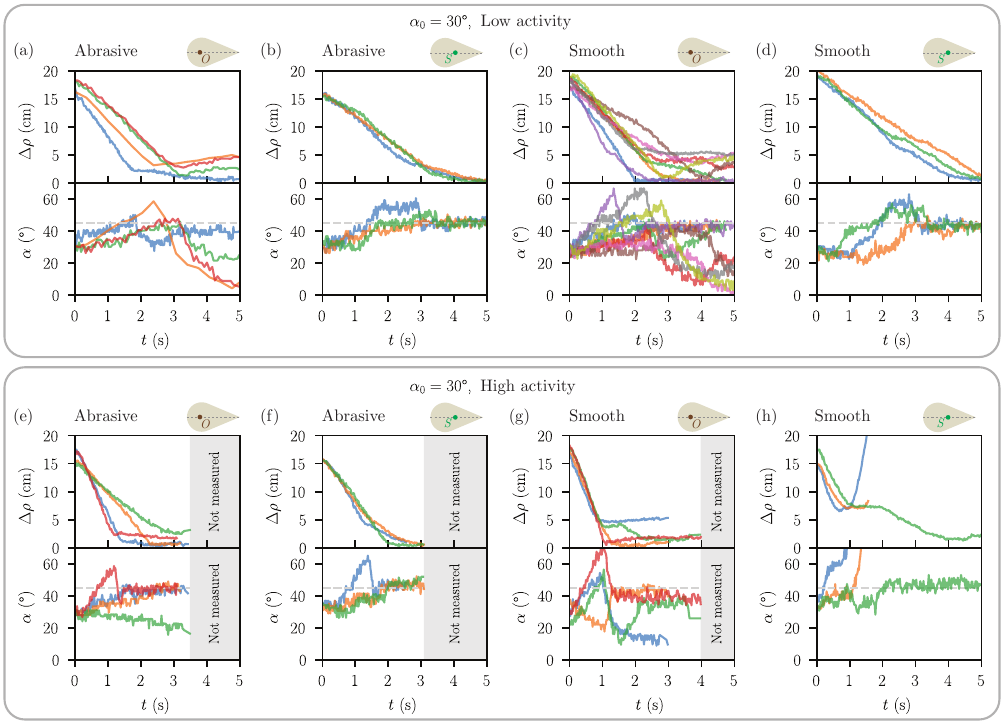}
    \caption{Kinematics of two-robot collisions for non-chiral self-propelled robots moving on a flat surface with the initial angle between the robots' propulsion vectors $\alpha_0=30^\circ$. (a)-(d) Time dependencies of the distance between the first and the second robots' noses $\Delta\rho$ and the angle between the robots' propulsion vectors $\alpha$. Solid lines with different colors in each panel correspond to different realizations of the collision experiment with the same initial angle between the robots moving at $\text{PWM}=10\%$. Individual parameters for the panels are the following: (a) $O$-robots with abrasive side surfaces, (b) $S$-robots with abrasive side surfaces, (c) $O$-robots with smooth side surfaces, and (d) $S$-robots with smooth side surfaces. The shaded gray areas in Panels (a), (c), and (d) denote correspond to robots reaching the region inaccessible for recording. The gray dashed line shows the angle $\alpha = 45^\circ$ characteristic of a ``heart'' cluster. (e)-(h) The same as Panels (a)-(d), but for the robots moving at $\text{PWM}=20\%$.}
    \label{figS:collision_angle_30}
\end{figure*}

\begin{figure*}[tbp]
    \centering
    \includegraphics[width=170mm]{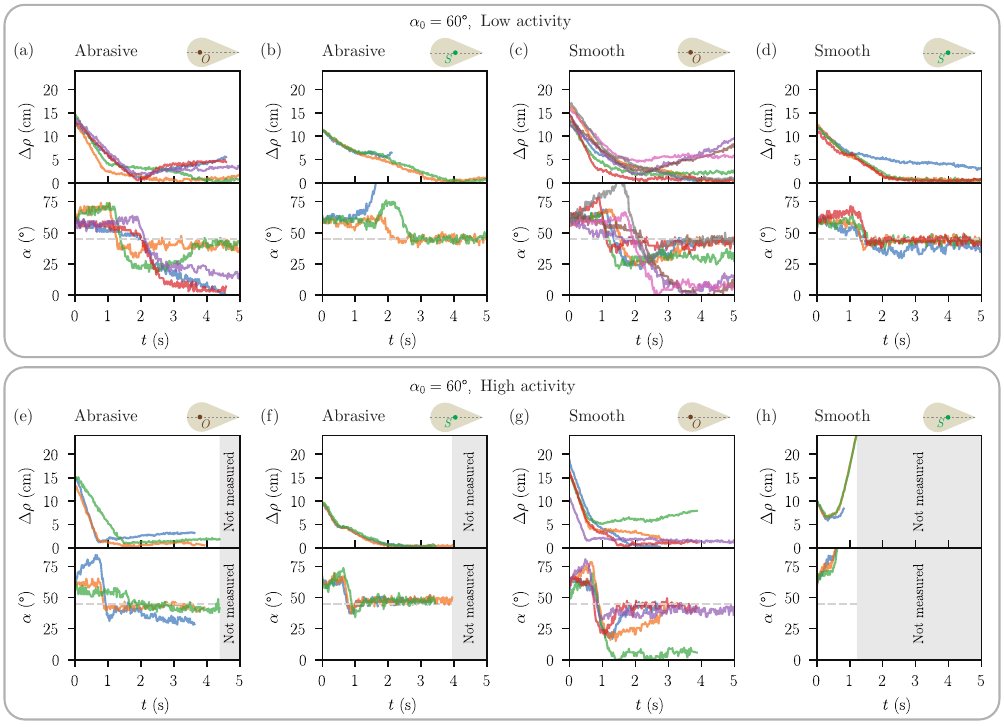}
    \caption{Kinematics of two-robot collisions for non-chiral self-propelled robots moving on a flat surface with the initial angle between the robots' propulsion vectors $\alpha_0=60^\circ$. (a)-(d) Time dependencies of the distance between the first and the second robots' noses $\Delta\rho$ and the angle between the robots' propulsion vectors $\alpha$. Solid lines with different colors in each panel correspond to different realizations of the collision experiment with the same initial angle between the robots moving at $\text{PWM}=10\%$. Individual parameters for the panels are the following: (a) $O$-robots with abrasive side surfaces, (b) $S$-robots with abrasive side surfaces, (c) $O$-robots with smooth side surfaces, and (d) $S$-robots with smooth side surfaces. The shaded gray areas in Panels (a), (c), and (d) denote correspond to robots reaching the region inaccessible for recording. The gray dashed line shows the angle $\alpha = 45^\circ$ characteristic of a ``heart'' cluster. (e)-(h) The same as Panels (a)-(d), but for the robots moving at $\text{PWM}=20\%$.}
    \label{figS:collision_angle_60}
\end{figure*}

The same trends are seen for the initial angle between the propulsion vectors of the robots $\alpha_0=30^\circ$, Fig.~\ref{figS:collision_angle_30}, and $\alpha_0=60^\circ$, Fig.~\ref{figS:collision_angle_60}. For $S$-robots, the formation of symmetric ``heart'' clusters is favored, as seen in Figs.~\ref{figS:collision_angle_30}(b),(d),(f),(h) and Figs.~\ref{figS:collision_angle_60}(b),(d),(f), while $O$-robots typically form asymmetric ``tandem'' clusters characterized by a non-vanishing distance between the first and the second robots' noses $\Delta\rho$ and approximately zero angle $\alpha$ corresponding to the aligned moving robots. The formation of such clusters corresponds to all curves in Figs.~\ref{figS:collision_angle_30}(c),(e), the red solid line in Fig.~\ref{figS:collision_angle_30}(g), the blue and orange solid lines in Figs.~\ref{figS:collision_angle_60}(a),(c), and the red, orange and green solid lines in Fig.~\ref{figS:collision_angle_60}(e).

Moreover, as the initial angle between the robot propulsion vectors $\alpha_0$ and the robot activity increase, robots are more likely to collide in a nose-to-side fashion and are unable to form a cluster, resulting in the scattering of robots. There are three such experiments for $\alpha_0=30^{\circ}$ corresponding to the orange solid lines in Figs.~\ref{figS:collision_angle_30}(a),(h) and the blue solid lines in Fig.~\ref{figS:collision_angle_30}(h). For $\alpha_0=60^{\circ}$, there are eight such experiments: the green solid line in Figs.~\ref{figS:collision_angle_60}(a),(f), the blue solid lines in Figs.~\ref{figS:collision_angle_60}(b),(e),(g), the blue, orange, and green solid lines in Fig.~\ref{figS:collision_angle_60}(h). The reason for the growing proportion of experiments that result in scattering is the increasing difficulty of engineering the collision between robots at higher robot activity and larger angles $\alpha_0$.

\section{Classification of different several-robot clusters}
\label{sec:Clusters}

\begin{figure*}[tbp]
    \centering
    \includegraphics[width=170mm]{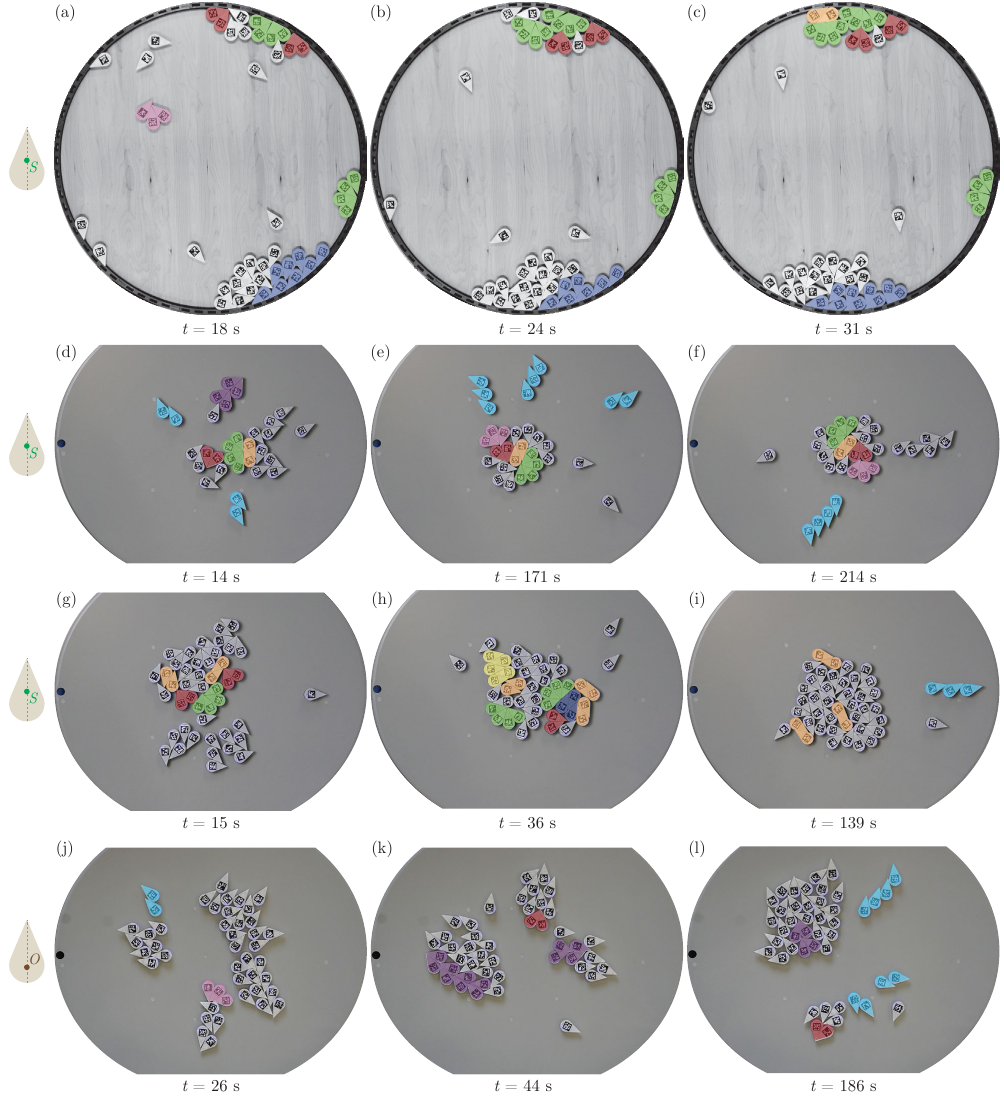}
    \caption{Examples of different $n$-robot clusters shaded with different colors (a)-(c) at the flat surface enclosed by a circular barrier with the diameter $D=90$~cm and (d)-(l) in the swarms of robots placed in a parabolic potential. (a)-(c) Snapshots of the swarm motion for $N=46$ $S$-robots at the timestamps (a) $t=18$~s, (b) $t=24$~s, and (c) $t=31$~s. (d)-(f) Snapshots of the swarm motion for $N=30$ $S$-robots at the timestamps (d) $t=14$~s, (e) $t=171$~s, and (f) $t=214$~s. (g)-(i) Snapshots of the swarm motion for $N=45$ $S$-robots at the timestamps (g) $t=15$~s, (h) $t=36$~s, and (i) $t=139$~s. (j)-(l) Snapshots of the swarm motion for $N=45$ $O$-robots at the timestamps (j) $t=26$~s, (k) $t=44$~s, and (l) $t=186$~s.}
    \label{figS:Clusters}
\end{figure*}

Along with complete micelles consisting of eight robots orienting their noses to the common point, a variety of clusters consisting of $n \ge 2$ robots can be formed. In Figure~\ref{figS:Clusters}, we introduce the following classification of various clusters that are either stable (i.e., can propagate with their shape conserved) or can exist in contact with system boundaries or other clusters:

\begin{enumerate}
   
   \item ``Heart-shaped'' clusters ($n=2$). The most common cluster that has the form of a pair of robots whose cusps point towards the common point [red-shaded clusters in Figs.~\ref{figS:Clusters}(a)-(h),(j),(l)].
   
   \item ``Yin-yang'' clusters ($n=2$). In such a cluster, two robots touching with their side surfaces point their noses towards each other's backs [orange-shaded clusters in Figs.~\ref{figS:Clusters}(c)-(i)]. 
   
   \item ``Incomplete 3/8 micelles'' ($n=3$) that are formed by $n=3$ touching robots [pink-shaded clusters in Figs.~\ref{figS:Clusters}(a),(e)-(f),(j)].
   
   \item ``Semi-micelles'' ($n=4$) formed by $n=4$ touching robots [green-shaded clusters in Figs.~\ref{figS:Clusters}(a)-(h)].

   \item ``Tandem'' ($n=2$) and ``chain'' ($n > 2$) clusters formed by robots that touch the side surfaces of neighboring robots with their noses [cyan-shaded clusters in Figs.~\ref{figS:Clusters}(d)-(f),(i)-(j)].
   
   \item ``Double-chain'' clusters ($n \ge 4$, $n \in {\rm even}$) that have the form of linear chains of ``heart-shaped'' clusters of $n=2$ robots [purple-shaded clusters in Figs.~\ref{figS:Clusters}(d),(j)-(l)].

   \item ``Boundary-localized double-chain'' clusters ($n \ge 5$). These structures can be considered as ``double-chain'' clusters that are deformed by a curved boundary such that robots at the barrier form a chain with their noses touching side surfaces of neighboring robots, and additional robots can join the cluster with their noses touching the contact point of two adjacent robots in the linear chain [blue-shaded clusters in Figs.~\ref{figS:Clusters}(a)-(c)].

   \item ``V-shaped'' clusters ($n = 3$) representing a symmetric combination of a ``yin-yang'' cluster with an additional robot, which can also be considered as two ``yin-yang'' clusters merged through a common central robot [the dark blue-shaded cluster in Fig.~\ref{figS:Clusters}(h)].

   \item ``Densely packed'' clusters ($n \ge 4$) formed by sequentially merged ``yin-yang'' clusters [the yellow-shaded cluster in Fig.~\ref{figS:Clusters}(h)].
   
\end{enumerate}

It is important to note that some arrangements of robots can be decomposed into combinations of touching clusters in several ways. Some examples of clearly pronounced clusters are highlighted in Fig.~\ref{figS:Clusters}. Moreover, it is seen that common general patterns in cluster occurrence and stability are observed for swarms of $S$-robots on a flat surface, Figs.~\ref{figS:Clusters}(a)-(c), and in the parabolic potential, Figs.~\ref{figS:Clusters}(d)-(i). For example, unstable ``yin-yang'' clusters [the orange clusters in Figs.~\ref{figS:Clusters}(c)-(i)] appear in both cases, as well as stable ``heart-shaped'' clusters. Additional examples of stable structures include ``double-chain'' clusters that can move along elliptic trajectories in the parabolic dish in a drift-like manner, an ``incomplete $3/8$ micelle'' of robots moving in contact until a collision with other structures [the pink clusters in Figs.~\ref{figS:Clusters}(a),(e),(f),(j)]. At high system densities, Figs.~\ref{figS:Clusters}(g)-(i), several clusters of robots frequently merge into one large cluster that begins to rotate. 

In contrast, a system with $O$-robots demonstrates different cluster patterns, Fig.~\ref{figS:Clusters}(j)-(l). Similarly to the previous system, stable ``double-chain'' clusters and ``heart-shaped'' clusters exist. However, the cluster of three robots is less stable and appears less frequently, as well as a ``semi-micelle''. Instead, ``linear chain'' clusters and ``double-chain'' clusters are formed more frequently and include more robots.

\section{The effects of aggregation induced by parabolic potential}
\label{sec:Aggregtion}

As seen in Fig.~5 in the main text, the parabolic potential induces aggregation of robots. This aggregation demonstrates different properties for $O$-robots and $S$-robots. While $O$-robots condense into aligned linear structures directed outwards the potential vertex and located at some characteristic height [Fig.~5(a)], $S$-robots form dense structures around the potential vertex characterized by disordered orientations of individual robots [Fig.~5(b)]. Here, we discuss how this aggregation affects the formation of micelle-like clusters.

To proceed, we perform a set of additional experiments within which we first artificially assemble a micelle-like cluster in the vicinity of parabolic potential vertex for both $O$- and $S$-robots, then randomly add some robots around this cluster, and activate the robots' motion. The results are shown in Table~\ref{tabS:Potential_micelle_time}.

It is seen that for all configurations ($O$- and $S$-robots with low or high vibration activity and abrasive or smooth side surfaces), such artificially assembled micelles for $N=8$ (no robots around the micelle) appear very stable and exist for the entire observation time within the experiment, which is a few minutes. This demonstrates that the absence of micelles in the swarms of $O$-robots shown in Table~III and Fig.~6 is related not to the instability of such micelles, but to the suppression of the microscopic mechanisms responsible for their formation.

When we add additional robots, the situation changes and micelle-like clusters can now disassemble due to collisions with other robots. We start with the case of robots with smooth side surfaces. As seen in Table~\ref{tabS:Potential_micelle_time}, in systems of $N=15$ $O$-robots, these initially created micelle-like clusters demonstrate typical lifetimes of just several seconds either for low or high activity of robots. For $S$-robots, the situation is the same. However, a relatively stable micelle is observed in the system with high vibration activity. If the number of robots in the swarm is increased to $N=30$, $O$-robots demonstrate the same micelle lifetime of several seconds for low activity, but the micelle is present throughout the experiments for high activity. The reason behind such a behavior is the following. As seen in Fig.~5, $O$-robots align at some characteristic height. For higher activity, this level is also higher, and, as a result, the robots not involved in the formation of the micelle-like cluster are less likely to interact with it. For $S$-robots, the behavior is different: while at low activity a formation of four new micelle-like clusters is observed after the disassembly of the initial one, for high activity there were not any subsequent reassembly.

For robots with abrasive surfaces, the micelle-like clusters formed by $O$-robots do not disassemble for $N=15$. For $N=30$, the micelle quickly disassembles at low activity, but then reassembles due to collisions with other robots and is observed almost during the entire experiment ($120$~s) at high activity. Thus, increased friction between side surfaces rather increases micelle lifetimes for $O$-robots, as all clusters, including micelle-like ones, become more stable, and the robots that are not involved in the formation of a micelle-like cluster aggregate at some characteristic height, which lowers their interaction with this cluster. For $S$-robots, the micelle lifetime increases for $N=15$ and low vibration activity, but decreases for high vibration activity. For $N=30$ and low vibration activity, there are only two micelle-like clusters formed after the disassembly of the initially placed one, and their joint lifetime of $46$~s is considerably shorter than $81$~s corresponding to smooth side surfaces. Finally, for $N=30$ and high vibration activity, there is a single micelle-like cluster formed after the initial one, with their total lifetime being $10$~s that is slightly greater than in the case of smooth side surfaces, but is still by order of magnitude lower compared to the observation period.

To summarize, it is seen that for $S$-robots as well as $O$-robots, the effects of aggregation counteract the existence of micelle-like clusters. Since in $O$-robots such clusters do not form in a natural way, this can be shown only in the artificial setting analyzed in Table~\ref{tabS:Potential_micelle_time}. They either decompose in the first few seconds if a collision with a non-micellized robot occurs, or remain stable during the entire experiment as the non-micellized $O$-robots move away from the potential vertex and do not collide with the micelle anymore. For $S$-robots, micelle-like cluster lifetimes are likely to decrease with increasing number of robots in the system as well as with increasing friction between robot side surfaces, i.e., when the aggregation itself becomes more pronounced or the clusters formed by robots become more stable.

\setlength\tabcolsep{0.01cm}
\begin{table*}[tbp]
    \renewcommand{\arraystretch}{1.3}
    \centering
    \begin{tabularx}{\textwidth} { 
   >{\centering\arraybackslash}X
   >{\centering\arraybackslash}X
   >{\centering\arraybackslash}X >{\centering\arraybackslash}X >{\centering\arraybackslash}X >{\centering\arraybackslash}X >{\centering\arraybackslash}X >{\centering\arraybackslash}X >{\centering\arraybackslash}X >{\centering\arraybackslash}X >{\centering\arraybackslash}X >{\centering\arraybackslash}X >{\centering\arraybackslash}X }
        \hline\hline
        \multicolumn{12}{c}{Smooth lateral surfaces} 
        \\\hline
        \multicolumn{6}{c}{Low activity}
         & 
         \multicolumn{6}{c}{High activity}
         \\\hline
         \multicolumn{2}{c}{$N=8$}
         &
         \multicolumn{2}{c}{$N=15$}
         &
         \multicolumn{2}{c}{$N=30$}
         &
         \multicolumn{2}{c}{$N=8$}
         &
         \multicolumn{2}{c}{$N=15$}
         &
         \multicolumn{2}{c}{$N=30$}
         \\\hline
         $O$ & $S$ & $O$ & $S$ & $O$ & $S$ & $O$ & $S$ & $O$ & $S$ & $O$ & $S$
         \\\hline
         
         0--end & 0--end & 0--4~s & 0--3~s & 0--5~s & 0--2~s & 0--end & 0--end & 0--3~s & 0--53~s & 0--end & 0--2~s  
         \\
         (150~s) & (150~s) & &  &  & 15--24~s & (150~s) & (150~s) &  &  & (150~s) &
         \\
         & & &  &  & 32--53~s & &  &  &  & &
         \\
         & & &  &  & 97--112~s & &  &  &  & &
         \\
         & & &  &  & 284--318~s & &  &  &  & & 

         \\\hline\hline
         \multicolumn{12}{c}{Abrasive lateral surfaces} 
        \\\hline
        \multicolumn{6}{c}{Low activity}
         & 
         \multicolumn{6}{c}{High activity}
         \\\hline
         \multicolumn{2}{c}{$N=8$}
         &
         \multicolumn{2}{c}{$N=15$}
         &
         \multicolumn{2}{c}{$N=30$}
         &
         \multicolumn{2}{c}{$N=8$}
         &
         \multicolumn{2}{c}{$N=15$}
         &
         \multicolumn{2}{c}{$N=30$}
         \\\hline
         $O$ & $S$ & $O$ & $S$ & $O$ & $S$ & $O$ & $S$ & $O$ & $S$ & $O$ & $S$
         \\\hline
         
         0--end & 0--end & 0--end & 0--end & 0--3~s & 0--8~s & 0--end & 0--end & 0--end & 0--5~s & 0--2~s & 0--5~s
         \\
         (150~s) & (150~s) & (180~s) & (180~s) &  & 17--31~s & (150~s) & (150~s) & (120~s) &  & 6--end & 8--13~s
         \\
         & & &  &  & 56--80~s & & & & & (120~s) &
         
         \\\hline\hline
    \end{tabularx}
    \caption{Lifetime of manually assembled micelles in swarms of $N=8$, $N=15$, and $N=30$ self-propelled robots in a parabolic potential for $O$-robots and $S$-robots moving with low activity ($\text{PWM}=10\%$) and high activity ($\text{PWM}=20\%$), and two types of robots' lateral surfaces: abrasive and smooth. Two numbers in each cell correspond to the time of micelle formation ($t=0$~s corresponds to the micelle assembled manually) and the time of its decomposition. If more micelles spontaneously form during a particular experiment, the corresponding cell contains multiple time ranges. If the micelle did not decompose until the end of the experiment, ``end'' is used in the corresponding time range and the duration of the experiment is specified in brackets.}
    \label{tabS:Potential_micelle_time}
\end{table*}


\section{Comparison with micellization of surfactants}
\label{sec:Surfactants}

The reported formation of micelle-like clusters in robotic swarms shares some features with the usual micellization, e.g., a broken symmetry of the particles that orient themselves in a specific way producing a cluster with a rotational symmetry. However, there is a set of important dissimilarities. The latter include (i) the absence of hydrogen bonds, which are the main driving force of surfactant micellization; (ii) only one kind of particles present in our system, as opposed to a mixture of solvent and surfactant molecules, where micellization usually occurs; and (iii) our system being inherently non-equilibrium as the robots constantly convert the energy stored in their batteries to kinetic energy. It is therefore interesting whether a set of parameters allowing straightforward comparison with the usual micellization of surfactants can be introduced. One of the most important parameters in surfactant micellization is the critical micelle concentration (CMC), defined as the lowest concentration of the surfactant at which micelles start to form~\cite{1955_Phillips,2014_Nagarajan}. In this section, we speculate whether it is possible to define its analog in our system.

\begin{figure*}[tbp]
    \centering
    \includegraphics[width=170mm]{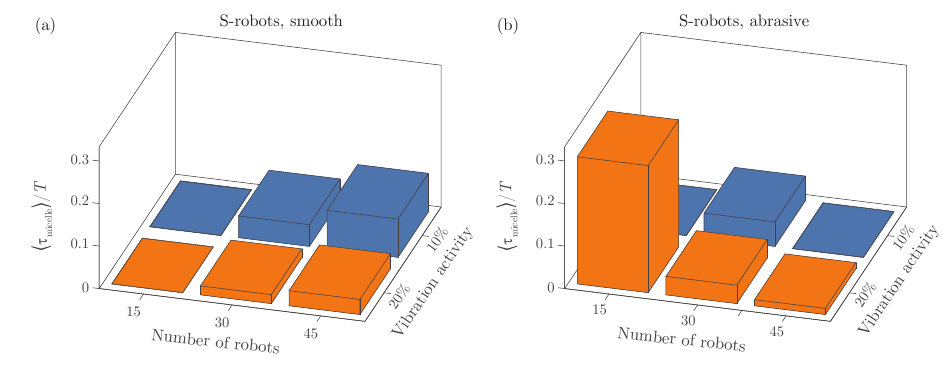}
    \caption{Ratio of the total time a micelle is observed in experiments to the total experiment duration at given set of parameters: the number of robots and their vibration activity. Panel (a) shows $S$-robots with smooth lateral surfaces; panel (b) shows $S$ robots covered with an abrasive. $O$-robots are not shown as they do not display formation of micelles.}
    \label{figS:ActivityDensityAxes}
\end{figure*}

A conventional theoretical approach of obtaining the analytical expressions for the free energy of micellization, consists in considering micellization as a reversible chemical reaction
\begin{equation*}
    M\mathrm{S} \leftrightharpoons \mathrm{S}_M,
\end{equation*}
where $M$ is the number of S (placeholder for the surfactant) molecules in a single micelle, while $\mathrm{S}_M$ denotes a micelle. Using the expression for the equilibrium constant of this reaction, one can derive the free energy of micellization~\cite{1955_Phillips,2014_Nagarajan}, which also allows to obtain the expression for the temperature dependence of CMC. As is known from Le Chatelier's principle, increasing the concentration of one of the species participating in the reaction shifts the equilibrium point to the side that would counter that change in concentration. In micellization of surfactants, this leads to an increasing dependency of the micelle concentration on the overall surfactant concentration, starting from the CMC point where the micelles first appear. Therefore, if we were to apply the same theoretical approach to our system, ignoring for once that it is not in equilibrium, the formation of micelle-like clusters of robots would be expected to increase with the concentration of particles playing the role of a surfactant.

It is, however, not possible to define an analog of the surfactant concentration for our system in a rigorous manner, because it only has particles of one kind, as opposed to surfactant and solvent molecules. A simple approach would be to consider the robots as a surfactant and the surrounding space as a solvent. Despite the apparent lack of physical sense, this allows one to define the surfactant concentration as the robot filling density, i.\,e., the ratio of the area filled by robots to the entire area enclosed by the boundary (for experiments on the flat surface) or the area of a paraboloid up to the greatest height robots can reach (for experiments in the parabolic potential). The micelle concentration would then be defined as the surface covered by complete micelles divided by the area enclosed by the boundary.

With this definition of concentration, the micelle count would be expected to increase with the number of robots. To examine whether this is true, we consider the fraction of time a micelle is observed according to Table~III from the main text with respect to the total time of the experiments (900~s) in a given configuration, which is proportional to micelle concentration. The results are shown in Fig.~\ref{figS:ActivityDensityAxes}. For smooth robots, the micelle concentration indeed increases with the number of robots at both vibration activities; see Fig.~\ref{figS:ActivityDensityAxes}(a). However, the situation changes qualitatively for abrasive robots, as seen from Fig.~\ref{figS:ActivityDensityAxes}(b): a decreasing dependence is observed instead for high vibration activity, and a non-monotonic dependence for low vibration activity. Moreover, further increase in the number of robots, which would lead to a further increase in micellization in surfactant systems, will, instead, prevent it completely in our system due to the jamming transition~\cite{2019_Barois}.

It is also known that micelles in surfactant systems only form at temperatures exceeding the Krafft temperature. Therefore, another interesting question is whether the vibration activity of robots can be considered as an analog of temperature in the context of micellization. There are a number of active matter systems that were demonstrated to have an effective temperature of some kind~\cite{2008_Loi,2015_Takatori,2022_Sanjay,2022_Boudet,2024_Caprini}. However, it is not clear to what degree this is applicable to our system, at least without a detailed theoretical analysis. Assuming that the effective temperature is proportional to the vibration activity, let us consider Fig.~\ref{figS:ActivityDensityAxes} as an analog of a phase diagram for micellization. It is seen that for abrasive robots micelle-like clusters do not form for low vibration activity at low and at high densities; see Fig.~\ref{figS:ActivityDensityAxes}(b). However, for smooth robots lower vibration activity corresponds to a somewhat more effective micellization instead, as is seen from Fig.~\ref{figS:ActivityDensityAxes}(a). Therefore, it is unlikely that our system could exhibit an analog of the Krafft point.


%